\documentclass[letterpaper,12pt,peerreviewca,draftcls]{IEEEtran}
\usepackage{csm16}
\usepackage[margin=1in]{geometry}
\usepackage[nolists,nomarkers,tablesfirst]{endfloat}
\usepackage{amsmath,amssymb,amsfonts}
\usepackage{algorithmic}
\usepackage{textcomp}
\usepackage{gensymb}
\usepackage{comment}
\usepackage{mathtools}
\usepackage{url}
\usepackage{graphicx,xcolor}
\usepackage{verbatim}
\usepackage[left,pagewise]{lineno} 
\usepackage[english]{babel} 
\usepackage{blindtext}
\graphicspath{{Figures/}}
\usepackage[compress]{cite}
\usepackage{physics}
\usepackage{wrapfig}

\newif\ifPDF \ifx\pdfoutput\undefined\PDFfalse \else\ifnum\pdfoutput > 0\PDFtrue \else\PDFfalse \fi \fi
\ifPDF 
\usepackage[pdftex, plainpages = false, colorlinks=true, linkcolor=black, citecolor = green!50!blue, urlcolor = blue, filecolor=black, pagebackref=false, hypertexnames=false,  pdfpagelabels ]{hyperref}
\fi

\newcommand{\Ind}{$\dv*[n]{I}{V}$} 
\newcommand{\Ifd}{$\dv*{I}{V}$}
\newcommand{\Isd}{$\dv*[2]{I}{V}$}
\newcommand{\Itd}{$\dv*[3]{I}{V}$}

\title{Model-based Control of the Scanning Tunneling Microscope\\
\Large{Enabling New Modes of Imaging, Spectroscopy and Lithography}}

\author{Hamed Alemansour, S. O. Reza Moheimani\\
        Corresponding Author: S. O. R. Moheimani (reza.moheimani@utdallas.edu)\\
	\today}

\begin{document}
\maketitle
\CSMsetup
\modulolinenumbers[2] % added for CSMAG only

%%%%% Section 00 %%%%%%%%%%%%%%%%%%%%%%%%%%%%%
The principle of scanning tunneling microscopy is based on the quantum mechanical phenomenon of “tunneling”, whereby the wavelike properties of electrons allow them to “tunnel” beyond the surface of a solid into regions of space that are forbidden under the rules of classical physics \cite{wolf_2012}. This phenomenon is used in the scanning tunneling microscope (STM), where electrons tunnel from the apex of a sharp tip to a conducting surface held at a different potential. The probability of electrons tunneling decreases exponentially as the distance between the two surfaces increases \cite{Book_Voigtlaender}. Hence, electrons tunnel mostly from the very last atom on the tip apex to the surface. The STM makes use of this extreme sensitivity to distance. A positioner brings the sharp tip of a tungsten probe to a few angstroms distance from the sample surface. A bias voltage is applied between the probe tip and the surface, causing electrons to tunnel across the gap.

\begin{figure}[!htb]
    \centering
	\includegraphics[width=1.0\linewidth]{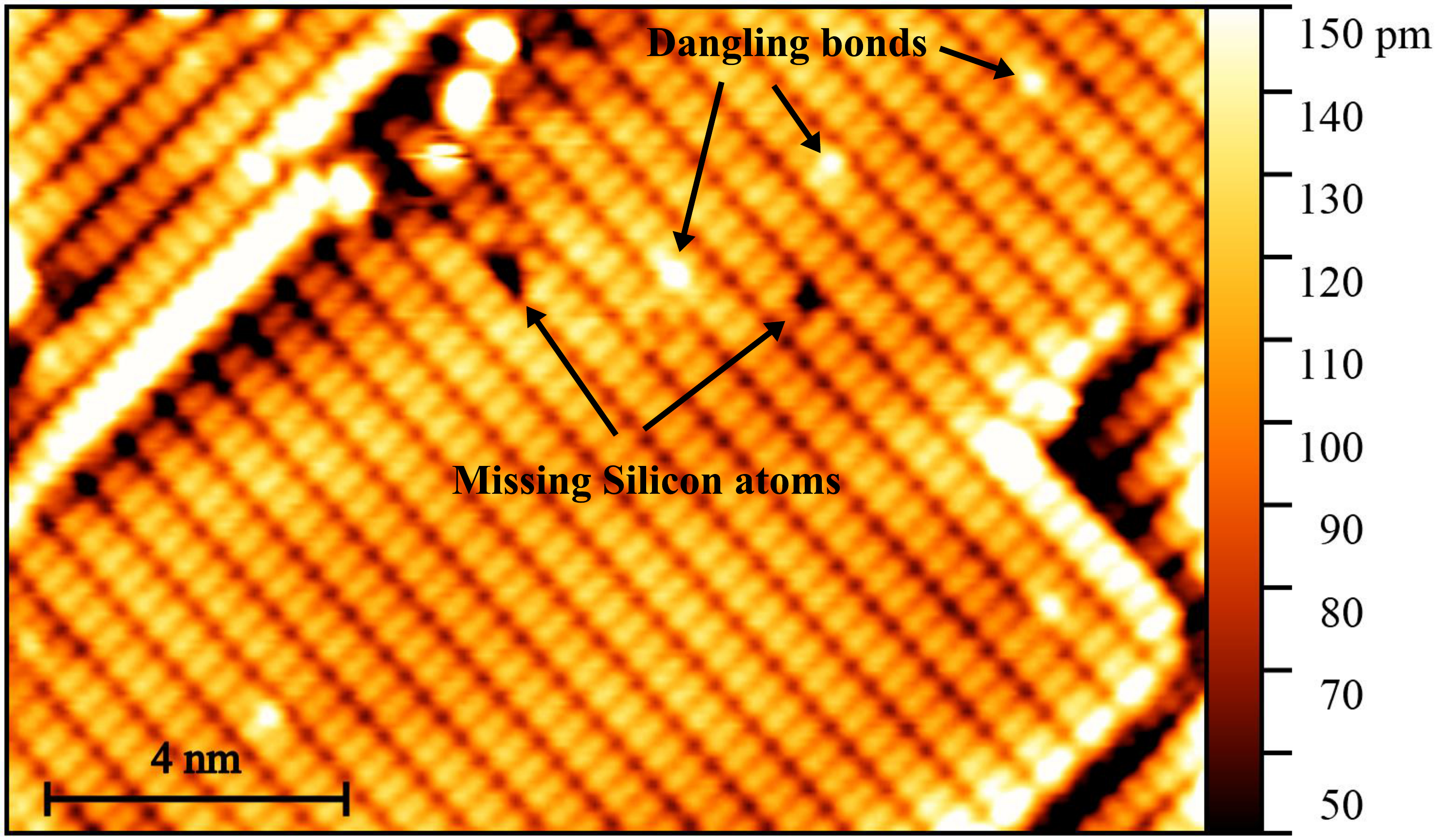}
	\caption{H-terminated silicon surface. The width of each distinct row (dimer row) is two silicon atoms. Direction of each dimer row changes by 90\degree{} when it meets a step edge. In this image, step edges progressively move down from the upper-left to the bottom-right corner. Dangling bonds (missing hydrogen atoms) and missing silicon atoms appear as the bright spots and dark areas, respectively.}
	\label{fig:Topography_SamExam}
\end{figure}

An STM  can function in several modes. In constant-current imaging mode, a controller adjusts the tip height to keep the tunneling current constant as the probe is scanned over the surface \cite{Farid_TCST}. Variations in the control signal are registered and processed to provide a topographical image of the surface. The STM image of a hydrogen-terminated \mbox{Si(100)-2$\times$1} surface is shown in Figure~\ref{fig:Topography_SamExam}. When the tip approaches a protrusion, the tunneling current increases in response to the reduction in the tip-sample distance. Consequently, the controller retracts the tip to recover the setpoint current. By plotting the controller output as a function of the in-plane position of the tip, a 3D map of the surface topography can be constructed.

Shortly after the initial demonstration of the STM, it was realized that the control effort does not reflect the true topography of a surface \cite{STS_Feenstra}.  Variations in electronic properties of the surface may result in distinct atoms, with identical heights, appearing differently on the obtained surface topography. A higher number of electrons tunnels through an atom with a higher electrical conductivity (that is, a lower barrier height), and the tip has to be retracted to maintain the setpoint current. Thus, local electronic properties do affect the tip height, resulting in an STM image that is a mixture of surface topographic and electronic features. In the STM image of an H-passivated Si surface in Figure~\ref{fig:Topography_SamExam}, bright dots represent missing hydrogen atoms on the surface, known as silicon dangling bonds. A true topography of the surface would show troughs at those spots. However, because of the higher electrical conductivity of silicon dangling bonds, they appear as protruding features on the constant-current topography image. Scanning tunneling spectroscopy (STS) was introduced soon after it was realized that an STM image is indeed a mixture of topographic and electronic features. The STS has opened new avenues to investigating the electronic structure of materials \cite{STS_Feenstra}. There are several distinctly different STS methods that will be discussed later in this article.

More recently, the STM  has been proposed as a lithography tool for atomically precise fabrication of micro-electronic devices \cite{Walsh_2009}. The possibility of performing both atomic resolution imaging and manipulation with the same tool distinguishes the STM from other scanning probe microscopes. Distinct advantages of a hydrogen-terminated silicon surface make it an ideal candidate  as a substrate for templating nanostructures with the STM \cite{Walsh_2009}. This surface is a chemically passivated Si substrate, where the surface Si atoms are bonded to H atoms. Hydrogen atoms on the top layer protect the highly reactive Si dangling bonds from forming further chemical reactions \cite{Lyding_1994, Lyding_Abeln_1994}. Therefore, this hydrogen layer can be considered as a resist. The STM tip can then be used to selectively remove H atoms by injecting current into the Si-H bonds at select locations. The reactivity and binding chemistry of  clean Si  allows for an assortment of molecules  to form chemical bonds with the resulting patterned reactive Si dangling bonds \cite{achal_2018_lithography, Butcher_2000_lithography, Adams_1996_lithography}. Devices fabricated using this method can be integrated with conventional micro-electronic devices and systems in existing semiconductor foundries \cite{Walsh_2009, Buch_2015_doping, Goh_2016_doping}.

An STM's functionality is highly dependent on the performance of its closed-loop feedback control system. The control system adjusts the tip position, relative to the surface (that is, the tip height in  imaging, spectroscopy, and lithography modes). Thus, it plays an important role in achieving the ultimate goal of fast, precise, and reliable STM operation. A poorly performing control system may result in the STM tip  crashing  into the surface, which tends to happen frequently in scanning tunneling microscopy resulting in many hours of lost work on a sample.  The performance of the feedback control system also affects the functionality of the STM when used for spectroscopy. Thus, a high-performance control system can enhance many aspects of STM operation, such as the tip lifetime, imaging speed, lithography precision,  and the quality of spectroscopic measurements. It also increases the probability that the system performs its designated functions (imaging, spectroscopy, and lithography) accurately and repeatedly without failure for a specified time period.

Despite its immense importance, little attention has been paid to the STM's control system since its invention in early 1980s. In this article, we study the effect of STM feedback control system on imaging, spectroscopy, and lithography performed with this instrument. We discuss various components of the STM feedback loop and their roles in the operation of the device. We propose experimental and analytical methods to obtain stability regions of an STM feedback control system. We describe novel control design methods to significantly improve the quality and operational speed of scanning tunneling spectroscopy. Finally, we discuss hydrogen depassivation lithography modes along with recently proposed control design methods to improve their precision.
%%%%%%%%%%%%%%%%%%%%%%%%%%%%%%%%%%%%%%%%%%%%%%%%%%

%%%%%%%%%%% Section01 %%%%%%%%%%%%%%%%%%%%%%%%%%%%%
\section{The scanning tunneling microscope system components}
\begin{figure}[!htb]
    \centering
	\includegraphics[width=.8\linewidth]{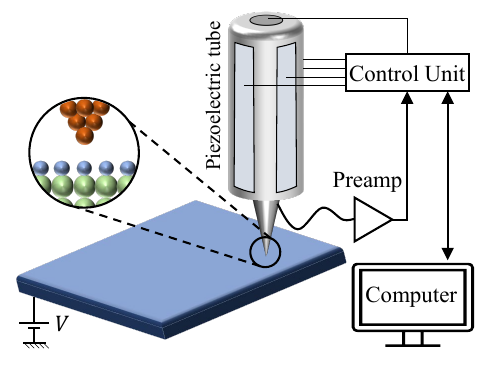}
	\caption{Schematics of an STM. A bias voltage V is applied between the tip and the sample that results in electrons tunneling from the tip to the surface  through the gap when the distance between the two objects is a few {\AA}ngstr{\"o}m. In this configuration, the tip is grounded. The current is regulated by a feedback control system that adjusts the tip-sample distance. The tip is moved in xyz-directions by a piezoelectric tube scanners.}
	\label{fig:STM_schematic}
\end{figure}

Various components of a typical STM are shown schematically in Figure~\ref{fig:STM_schematic}. A conductive probe with a sharp tip and a sample are brought into close proximity of each other, while a bias voltage is applied between them (for details, see ``Automated Approach of the Tip to the Sample Surface"). A tunneling current is established when the distance between the tip and the surface is approximately a nanometer. This current is amplified by a preamplifier (a transimpedance amplifier) and converted to a voltage that is measurable by the STM control unit. The control system produces appropriate command voltages to move the STM's scanner in x, y, and z directions. These voltages are amplified by a high-voltage amplifier and applied to the scanner.

\subsection{Piezoelectric Tube Scanner}
\begin{figure}[!htb]
    \centering
	\includegraphics[width=0.8\linewidth]{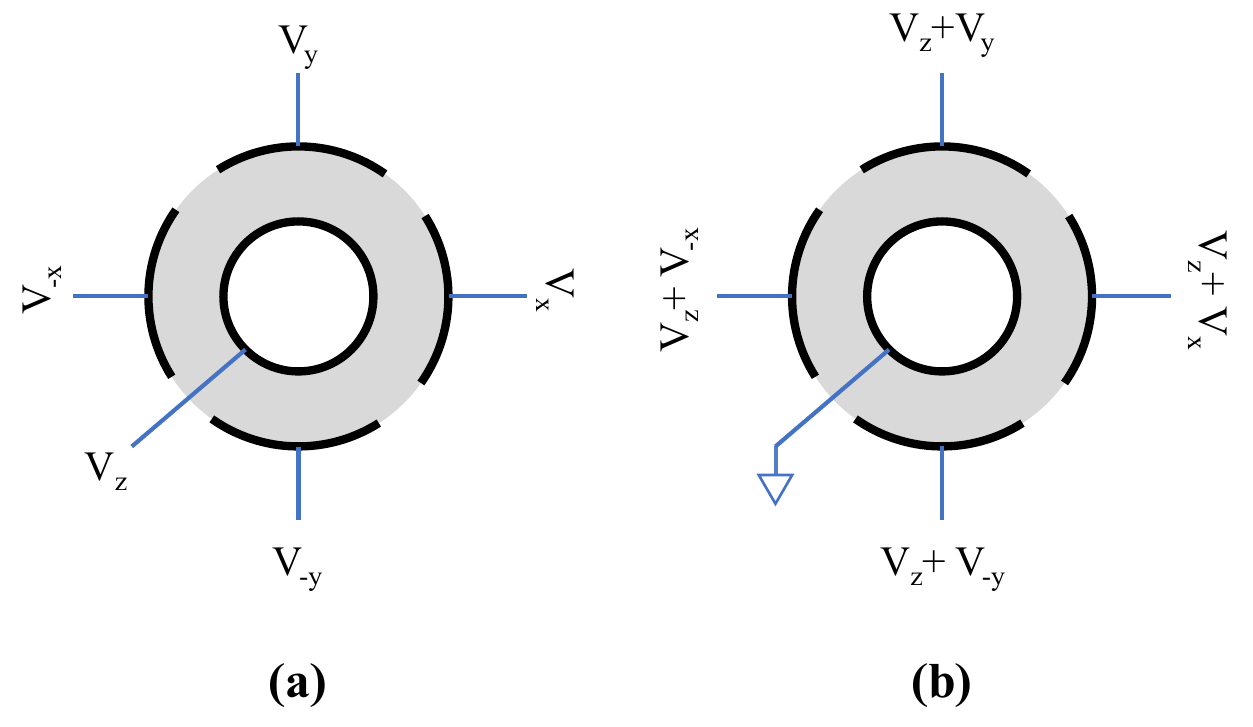}
	\caption{Two possible configurations for applying control voltages to a piezoelectric tube actuator. (a) The tip-sample height is controlled by applying the z actuation voltage to the inner electrode of the piezoelectric tube. (b) The z  displacement control voltage is mixed with the xy in-plane voltages and applied to the external electrodes.
	\label{fig:piezo}}
\end{figure}

The piezoelectric tube scanner is the most widely used method of actuation in STMs, owing to their outstanding displacement resolution. It is  easy to manufacture, and its integration  into a microscope is straightforward. It features a thin cylinder of radially poled piezoceramic material with an internal electrode and four quadrant external electrodes. Applying a voltage to one of the external electrodes results in a vertical expansion or contraction and, consequently, a lateral deflection of the tube. The tube is moved in x-direction by applying opposite polarity voltages, $V_{x}$ and $V_{-x}$, to the opposite quadrants along the x-axis. To produce motion along the y-axis, $V_{y}$ and $V_{-y}$ are applied to the remaining two electrodes. The high length-to-diameter ratio of a piezoelectric tube results in a large lateral displacement range. However, it also induces low mechanical resonance frequencies and limits the scanning bandwidth. There are two possible ways of actuating the tube along the z-direction. In the first approach, the z actuation signal is directly applied to the inner tube, as shown in Figure~\ref{fig:piezo}(a). Alternatively, the out of plane displacement is achieved by grounding the inner electrode and adding the z actuation voltage to the four quadrant external electrodes; see Figure~\ref{fig:piezo}(b). 

\begin{figure}[!htb]
    \centering
	\includegraphics[width=0.6\linewidth]{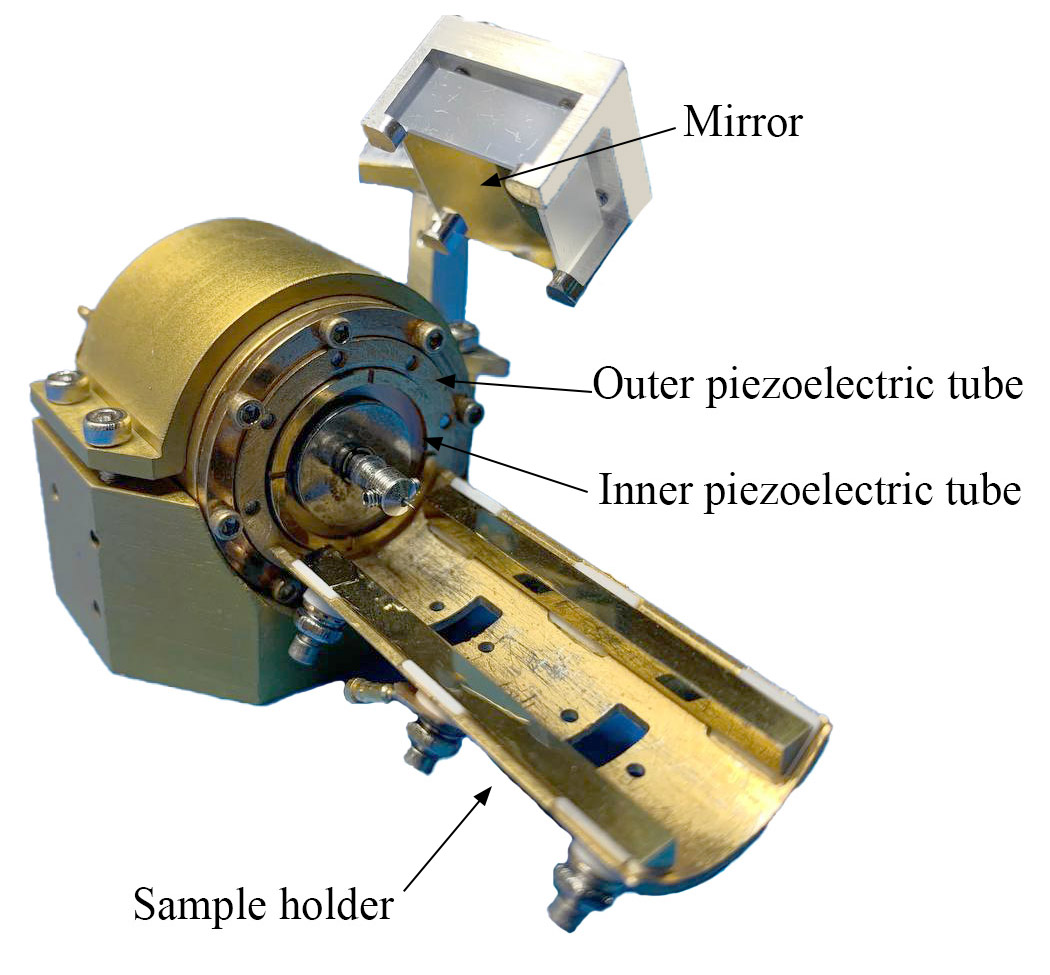}
	\caption{The Lyding scanner used in our STM. It consists of an outer piezoelectric tube scanner for  coarse positioning and an inner piezoelectric tube scanner for fine positioning the tip over a sample.}
	\label{fig:Lyding_scanner}
\end{figure}

A home-built Lyding-style~\cite{LydingScanner} ultra-high vacuum STM is used in the experiments reported throughout this article. The scanner is made of two concentric piezoelectric tubes as shown in Figure~\ref{fig:Lyding_scanner}. The inner tube, finely positions the tip relative to the sample and is also used for raster scanning the surface. The outer tube is used for coarse positioning and moves the sample toward the tip using a stick-slip mechanism.

\subsection{Current Amplifier}
\label{sec:preamp}
The tunneling current is only few nanoamperes and needs to be amplified before it is sent to the analog to digital converter (ADC). Therefore, a transimpedance amplifier (commonly known as the preamplifier) is an essential component of an STM. A preamplifier has a  high input impedance and low output impedance that enables converting small tunneling currents to measurable voltages. The preamplifier gain and its bandwidth have an inverse relationship. Thus, increasing the amplifier gain results in a lower bandwidth and {\it vice versa}. The amplification gain of a typical preamplifier is usually in the range of \mbox{$10^7$\,V/A} to \mbox{$10^9$\,V/A}. For an ADC with the input voltage range of \mbox{-10\,V} to \mbox{10\,V}, this limits the maximum allowable tunneling current to \mbox{10\,nA} to \mbox{1000\,nA}. Noise can drastically affect the small tunneling current measurements. To mitigate the noise effect, lowpass and notch filters can be incorporated in the feedback loop to filter out high frequency noise and disturbances with persisting frequency components.

\subsection{Lock-in Amplifier}
\label{subsec:demodulator}
STS relies on measuring tunneling electrons as a function of energy by varying the tip-sample voltage. This is typically performed by measuring differential conductance, \Ifd{}. Differential conductance is acquired by adding a small modulation voltage to the sample bias voltage and extracting the in-phase component of current via a demodulator. The STM does not require a demodulator for conventional imaging. However, STS methods rely on this technology. An STM can be equipped with a demodulator to obtain \Ifd{} image of a surface along with the topography.

A demodulator estimates amplitude and phase of a signal in presence of noise and/or additional frequency components. A periodic signal with the frequency of $\omega_c$ can be presented by a Fourier series as
\begin{equation} \label{eq:lockin_siganlofinterest}
    y(t) = a_0 + \sum_{n=1}^{\infty} {y_n(t)}
    =a_0 + \sum_{n=1}^{\infty} {a_n sin(\underbrace{n\omega_c}_{\omega_n} t + \phi_n)}
\end{equation}

\noindent where $a_n$ and $\phi_n$ are amplitude and phase of the signal at the frequency of $\omega_n$. At a frequency of interest $\omega_i=i \omega_c$, $y_i$ can be written as the sum of a quadrature and an in-phase component by applying trigonometric identities:
\begin{equation} \label{eq:yn_inphase_quadrature}
    y_i = \underbrace{a_i sin(\phi_i)}_{\mathclap{\text{quadrature component}}} cos(\omega_i t) + 
	\underbrace{a_i cos(\phi_i)}_{\mathclap{\text{in-phase component}}}sin(\omega_i t) = c_i X_i
\end{equation}

\noindent where
\begin{align}  \label{eq:yn_matrixform}
	c_i &= [cos(\omega_i t) \quad sin(\omega_i t)]\\
	X_i &= [x_{2i-1} \quad x_{2i}]^T =
	     [a_i\,sin(\phi_i) \quad a_i\,cos(\phi_i)]^T.
\end{align}

\noindent Then, amplitude and phase of interest are calculated as
\begin{equation}  \label{eq:lockin_amp_phase}
	a_i = \sqrt{{x^2_{2i-1}}+{x^2_{2i}}},\quad \phi_i = arctan\left( \frac{x_{2i-1}}{x_{2i}} \right).
\end{equation}

There are two demodulator classes  that could be used to extract amplitude and phase of a sinusoidal signal contaminated by noise. They are typically constructed based on methods that utilize rectification or a reference oscillator signal. RMS-to-DC conversion \cite{Kitchin_1986_RMStoDC}, peak hold \cite{Ando_2001_PeakHold}, and peak detector \cite{Ragazzon_2016_PeakDetector} are among the methods in the rectification-based category. These demodulators are not robust against unwanted frequency components and therefore cannot be employed when the signal contains spurious frequencies \cite{ruppert_2017_demodulation, Harcombe_2020_demodulation}. Methods that are based on the lock-in amplification \cite{Cosens_1934_LockInAmp, Michels_1941}, coherent demodulation \cite{Kokavecz_2006_CoherentDemodulator, Abramovitch_2011}, Kalman filtering \cite{Ruppert_2016_Kalman, Kalman_1960}, and Lyapunov filtering \cite{Ragazzon_2018_Lyapunov} are  demodulation methods that depend on a reference oscillator signal to function. These synchronous demodulators can provide estimates at multiple frequencies, a feature that makes them particularly suitable for STS applications where distinct frequency components of the current signal contain different information about the surface.

The lock-in amplification method has been adopted as the industry-wide standard demodulation technique in commercial STMs. It is less sensitive to other frequency components and noise, owing to its narrow tracking bandwidth. This bandwidth limitation also constrains the  scan rate achievable with this method.  A lock-in amplifier operates by multiplying the input signal described by \eqref{eq:lockin_siganlofinterest} with sine and cosine functions. The frequency of these  functions are set to  the frequency at which amplitude and phase of the input signal are to be estimated ($\omega_i=i \omega_c$). Therefore, the following signals are generated during the mixing process:
\begin{equation} \label{eq:lockin_inphase}
   \begin{aligned} 
     y_I(t) &= y(t) \times sin(\omega_i t) \\
     &= \sum_{n=1, n\ne{i}}^{\infty} {\frac{1}{2} a_n cos((\omega_n-\omega_i)t + \phi_n)} \\
     &- \sum_{n=1, n\ne{i}}^{\infty} {\frac{1}{2} a_n cos((\omega_n +\omega_i)t+\phi_n)} \\
     &+ a_0 sin(\omega_i t) + \frac{1}{2} [\underbrace{a_i cos(\phi_i)}_{x_{2i}} - a_i cos(2\omega_i t +\phi_i)]
   \end{aligned} 
\end{equation}
\noindent and
\begin{equation} \label{eq:lockin_quadrature}
    \begin{aligned}
        y_Q(t) &= y(t) \times cos(\omega_i t) \\
        &= \sum_{n=1, n\ne{i}}^{\infty} {\frac{1}{2} a_n sin((\omega_n-\omega_i)t+\phi_n)} \\
        &+ \sum_{n=1, n\ne{i}}^{\infty} {\frac{1}{2} a_n sin((\omega_n +\omega_i)t+\phi_n)} \\
	&+ a_0 cos(\omega_i t) + \frac{1}{2} [\underbrace{a_i sin(\phi_i)}_{x_{2i-1}} + a_i sin(2\omega_i t +\phi_i)].
    \end{aligned}
\end{equation}

It can be seen from \eqref{eq:lockin_inphase} and \eqref{eq:lockin_quadrature} that in addition to the dc components of interest, $x_{2i-1}$ and $x_{2i}$, other mixing products are also generated at the integer multiples of the modulation frequency. These high-frequency components, and also noise, are attenuated by a lowpass filter (LPF) whose order and bandwidth determine the tracking performance of the lock-in amplifier.
%%%%%%%%%%%%%%%%%%%%%%%%%%%%%%%%%%%%%%%%%%%%%%%%%

%%%%%%%%%%% Sec02 %%%%%%%%%%%%%%%%%%%%%%%%%%%%%%%
\section{Tunneling Current Model}
\label{sec:Tunneling_model}

Tunneling current depends on the density of states of the tip $\rho_{t}$ and the sample $\rho_{s}$ (see ``What is Quantum Tunnelling?"). Modeling of tunneling current in an STM is usually based on Bardeen's approximation of tunneling, which was developed long before the invention of STM. Bardeen's approximation of tunneling current is  
\begin{equation}\label{eq:Tunneling}
	I_{\text{Bardeen}} = \dfrac{4\pi e}{\hbar} \int_{0}^{eV} \rho_{t}(\epsilon - eV) \rho_{s}(\epsilon) T(\epsilon, V, \delta) d\epsilon
\end{equation}
where $\hbar$ is the Planck constant, V is the bias voltage, $\delta$ is the barrier thickness, and $T(\epsilon, V, \delta)$ is the transmission coefficient \cite{Book_Voigtlaender}. The transmission coefficient can be approximated as an  exponential function of the barrier thickness and the square root of the barrier height, that is, 
\[T \propto exp(-k \delta \sqrt{\phi}) \]
where $k$ is a constant, and $\phi$ is the local barrier height (LBH). Bardeen’s theory is valid, even when there is a large bias across the barrier (as is required in tunneling current through semiconductors). In addition, Bardeen’s theory can account for atomic-resolution \cite{Gottlieb_2006}. With some simplifying assumptions and after amplification of the tunneling current by the preamplifier, \eqref{eq:Tunneling} can be written as \cite{STM_Binnig01, Farid_RSI, Tajaddodianfar_2017_Control}
\begin{equation} \label{eq:SimplifiedTunn}
\begin{aligned}
    I &= R\,\dfrac{4\pi e}{\hbar} \int_{0}^{eV} \rho_{t}(\epsilon - eV) \rho_{s}(\epsilon) T(\epsilon, V, \delta) d\epsilon \\
	&\approx RL(V, \rho_{t}, \rho_{s}).e^{-1.025\,\delta\sqrt{\phi}}\\
\end{aligned}
\end{equation}
\noindent where $R$ is the preamplifier gain. $L$ is a function of voltage and the local density of states (LDOS) of tip and sample. This nonlinear relationship can be linearized  by taking the natural logarithm of both sides of (\ref{eq:SimplifiedTunn}), that is,
\begin{equation} \label{eq:LogSimplifiedTunn}
	\ln{(I)}\approx \ln{(RL)}-1.025\,\delta\sqrt{\phi} .
\end{equation}
Thus,  for a constant $L$, (\ref{eq:LogSimplifiedTunn}) implies that $\ln{(I)}$ is a linear function of the tip-sample gap. In addition, (\ref{eq:LogSimplifiedTunn}) can be used to show that the LBH is proportional to the square of derivative of $\ln{(I)}$ with respect to $\delta$,  
\begin{equation} \label{eq:barrier_height}
	\phi\approx 0.952\left(\dv{(\ln I)}{\delta}\right)^2.
\end{equation}

\noindent The barrier height provides vital information on surface electronic properties, for example, the behavior of electronic devices \cite{kahn_2016}, surface states \cite{fischer_1993}, and molecular adsorption and coverage \cite{jiang_2017}. It depends on the physical properties of the tip as well as the sample's local properties beneath the tip apex. Therefore, it is a local variable and can change by variations in the tip or sample properties. STM can be used to estimate local variations of barrier height. Existing methods for LBH estimation  will be discussed later in the article.
%%%%%%%%%%%%%%%%%%%%%%%%%%%%%%%%%%%%%%%%%%%%%%%%%

%%%%%%%%%%% Sec03 %%%%%%%%%%%%%%%%%%%%%%%%%%%%%%%
\section{STM Imaging}
\label{sec:imaging}
The conventional modes of imaging in an STM are constant height and constant current modes. The latter is a more robust mode of imaging due to the use of a feedback control system that keeps the tunneling current constant as the tip moves over the surface. We recently proposed a new mode of imaging, the  constant differential conductance mode  \cite{Alemansour_fastSTS}, which we  describe with the two conventional STM methods, below.

\subsection{Constant Height Imaging Mode}
\label{sec:constant_height_imaging}
\begin{figure}
    \centering
	\includegraphics[width=.9\linewidth]{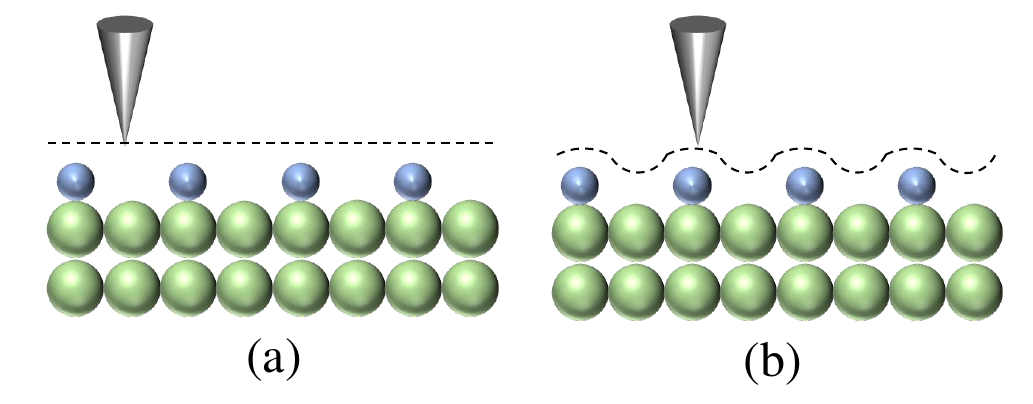}
	\caption{Two common imaging modes: (a) constant height, and (b) constant current. In constant height mode, the tip height does not change as it scans over the surface. In constant current mode, the tip height is adjusted by a closed-loop control system to keep the tunneling current constant.
	\label{fig:constant_current_height}}
\end{figure}

In constant height mode, depicted in Figure~\ref{fig:constant_current_height}{(a)}, the surface is scanned and the tunneling current is recorded while a constant voltage is applied to the z-axis of the piezoelectric tube scanner (that is, the tip position does not change with variations in surface topography); however, the tunneling current does. The natural logarithm of the tunneling current is then interpreted as the surface topography. In this mode, the high z-axis bandwidth of the system enables fast scans. However, the inevitable slope of the surface relative to the tip limits the scanning range to only a few nanometers. Performing a large area scan in constant height significantly increases the possibility of a tip-sample crash and is normally avoided. Furthermore, there is no possibility of protecting the tip from external disturbances and noise nor piezoelectric creep and drift in this mode. 

\subsection{Constant Current Imaging Mode}
\label{sec:constant_current_scanning}
\begin{figure}[!htb]
    \centering
	\includegraphics[width=1\linewidth]{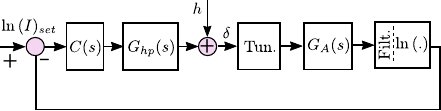}
	\caption{Block diagram of an STM feedback control system operating in  constant current imaging mode. Here, $C(s)$ and $G_A(s)$ represent the dynamics of the controller and the preamplifier, respectively. Also, $G_{hp}(s)$ represent $G_h(s)G_p(s)$, where $G_h(s)$ is the high-voltage amplifier and $G_p(s)$ is the piezo-electric tube actuator.
    \label{fig:block_diagram}}
\end{figure}

Constant current imaging mode is the most prevalent method of acquiring a surface topography with the STM. A block diagram of the z-axis closed-loop system is shown in Figure~\ref{fig:block_diagram}. In this mode, the tip-sample current is fed to a preamplifier with the amplification gain of $R$ and converted to a voltage. The preamplifer output is then sent to the STM control unit, where the natural logarithm of the current is calculated. Assuming that all parameters with the exception of $\delta$ are constant in \eqref{eq:LogSimplifiedTunn}, $\ln(I)$ would be a linear function of the tip-sample height and is therefore is used as the feedback signal. The error signal (that is, the difference between the setpoint and $\ln(I)$) is calculated and then is passed through the controller $C(s)$. The controller generates a command signal to minimize the error and keep the tip-sample current constant. The controller output is then amplified by a high-voltage amplifier $G_h(s)$ before it is applied to the STM's piezoelectric tube actuator $G_p(s)$. In Figure~\ref{fig:block_diagram}, $G_{hp}(s)$ represents the combined transfer function of the high-voltage amplifier and the piezoelectric tube actuator. The actuator moves the tip in the z-direction in response to the controller command. The controller command signal is then translated to displacement based on calibration coefficients of the actuator. This signal is plotted against the in-plane position of the tip to construct a topographic map of the surface. The STM tip path in the constant current imaging mode is schematically shown in Figure~\ref{fig:constant_current_height}{(b)}. The closed-loop operation of the STM in this mode enables scanning of larger areas, compared to the constant height imaging mode. Figure~\ref{fig:Topography_SamExam} shows the image of a H-terminated silicon surface obtained in this mode.

\subsection{Constant Differential Conductance (dI/dV) Imaging}
\label{sec:constant_dIdV_imaging}
\begin{figure}[!htb]
    \centering
	\includegraphics[width=1\linewidth]{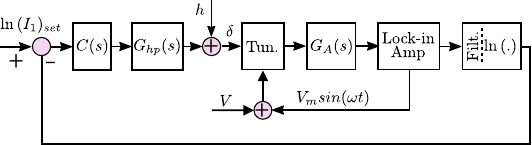}
	\caption{Constant differential conductance imaging mode. A modulation signal is applied to the sample, and the in-phase component of the tunneling current ($I_1$) is measured by a lock-in amplifier. The feedback loop is then closed on $\ln{I_1}$. For a small amplitude modulation signal, $I_1$ is proportional to \Ifd{}.
	\label{fig:BlockDiagram_dIdVfeedback} }
\end{figure}

The authors recently demonstrated that a topography image can be obtained in the constant differential conductance imaging mode \cite{Alemansour_fastSTS}. From \eqref{eq:SimplifiedTunn}, the natural logarithm of the first derivative of I with respect to V is
expressed as
\begin{equation} \label{eq:LogSimplifieddIdV}
	\ln{(\dv{I}{V})}\approx \ln{(R\dv{L}{V})}-1.025\,\delta\sqrt{\phi} .
\end{equation}
The natural logarithm of differential conductance $\ln(\dv*{I}{V})$ is linearly proportional to the tip-sample separation, $\delta$, assuming that all other parameters  are constant. This signal can be used as the measurement in a feedback loop to keep \Ifd{} constant. The control block diagram of an STM operating in this mode is shown in Figure~\ref{fig:BlockDiagram_dIdVfeedback}. A sinusoidal voltage $V_m\,sin(\omega t)$ is added to the dc bias voltage of the sample, with the tip electrically grounded. This results in high-frequency components in the current signal at harmonics of the fundamental frequency of the modulation voltage, due to the nonlinear current-voltage relationship. In-phase and  quadrature components of the current at the fundamental frequency are measured using a demodulator.
In the section on STS, we explain that for a small amplitude modulation voltage, the in-phase component of current at the fundamental frequency is proportional to \Ifd{}. This signal is sent to the STM control unit, and its natural logarithm is compared with the setpoint. The resulting error signal is then used by the controller to generate a control voltage that is applied to the z-axis of the piezoelectric tube. The tip height is then plotted as a function of the tip position to obtain an image of the surface topography. The PI controller design and the closed-loop system identification will be further elaborated on in the subsequent sections.

\begin{figure}[!htb]
	\noindent
	\centering
	\includegraphics[width=1\linewidth]{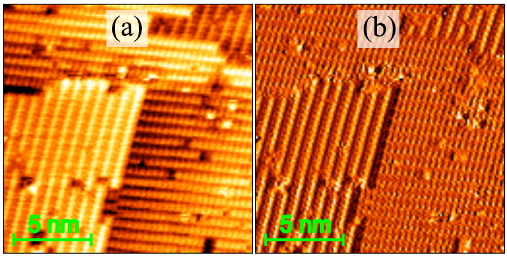}
	\caption{(a) Topography image, and (b) \Ifd{} (feedback signal) image of a hydrogen passivated silicon surface in constant differential conductance imaging mode. The controller adjusts the tip-sample height to keep \Ifd{} constant. Figure adapted from \cite{Alemansour_fastSTS}, with permission from AIP.
	\label{fig:dIdVimaging}}
\end{figure}

Figure~\ref{fig:dIdVimaging} shows an image of a hydrogen passivated silicon surface obtained in this mode \cite{Alemansour_fastSTS}. In this experiment, the bias voltage, modulation frequency, and modulation amplitude were \mbox{-\,2.5\,V}, \mbox{2\,kHz}, and \mbox{0.8\,V}, respectively.  The controller works to keep \Ifd{} constant at \mbox{0.125\,nA/V}. The topography image and image of the feedback signal (\Ifd{}) are shown in Figures~\ref{fig:dIdVimaging}{(a)} and \ref{fig:dIdVimaging}{(b)}, respectively.
%%%%%%%%%%%%%%%%%%%%%%%%%%%%%%%%%%%%%%%%%%%%%%%%%

%%%%%%%%%%% Sec04 %%%%%%%%%%%%%%%%%%%%%%%%%%%%%%%
\section{System Modeling and Proportional-Integral Controller Design}
\label{sec:system_modeling}
Proportional–integral (PI) control is universally used in commercial STMs. Many important characteristics of the STM (such as the maximum tip speed, noise, and closed-loop stability margins) are affected by the PI controller gains. Therefore, the controller design plays a crucial role in the efficient operation of the system. A low-bandwidth PI controller is more immune to sensor noise; however, it reduces the maximum achievable scan speed. On the other hand, a high-bandwidth feedback loop achieves a higher scan rate at the cost of lower stability margins. The stability region of the STM feedback control system was analytically obtained in \cite{Control_Oliva_1995, Control_Anguiano_1998} by mathematically modeling every component in the STM feedback loop. To build such a model requires accurate characterization of every component of the system. This approach is time consuming, difficult, and more likely to be prone to error. In this section, we discuss an alternative approach based on system identification and describe PI controller design for constant current and constant differential conductance imaging modes.

\subsection{System identification}
\label{subsec:constant_current_system_modeling}
\begin{figure}[!tbh]
    \centering
	\includegraphics[width=1\linewidth]{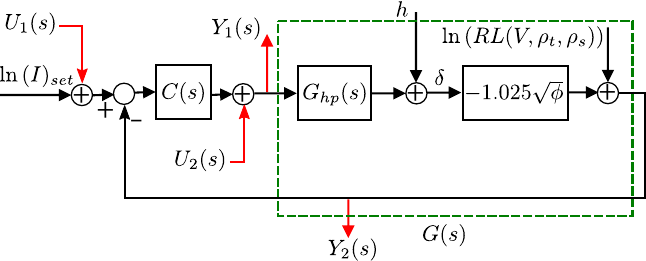}
	\caption{
                Block diagram of the STM z-axis control system in the constant current imaging mode. The four underlying transfer functions are represented as $G_{ij}(s)= {Y_i(s)}/{U_j(s)}$.}
	\label{fig:block_diagram_sysid}
\end{figure}

The simplified block diagram of an STM feedback loop operating in constant current imaging mode is shown in Figure~\ref{fig:block_diagram_sysid}. This block diagram is constructed based on the simplified tunneling current model \eqref{eq:LogSimplifiedTunn}. The open-loop z-axis model of the STM is represented by $G(s)$. It incorporates all of the components from the control command to the natural logarithm of current. A method was proposed in \cite{Farid_TCST} to obtain open-loop frequency response function (FRF) of $G(s)$ when the STM is operating in  closed-loop mode. The required experimental setup is shown in Figure~\ref{fig:block_diagram_sysid}. FRFs are obtained from inputs $U_1(s)$ and $U_2(s)$ to outputs $Y_1(s)$ and $Y_2(s)$. The open-loop transfer function $G(s)$ is obtained by dividing the FRFs at each frequency point according to
\begin{equation} \label{eq:closedloop_frf}
	G(s) = \frac{G_{21}(s)}{G_{11}(s)}= 
	\frac{G_{22}(s)}{G_{12}(s)}.
\end{equation}

The frequency response of $G(s)$ is obtained while the feedback loop is active and the tunneling current is established. The tip is also intended to remain idle relative to the sample during this process. As local properties of the sample (for example, LBH and LDS) and tip-sample height may change from one atom to another, the tip is placed where there are identical atoms in a relatively small area so that small drift of the piezoelectric tube actuator along the x- or y-axis would not affect the frequency response measurement. The frequency response functions were obtained experimentally with a frequency sweep from \mbox{100\,Hz} to \mbox{4500\,Hz}, while the tip-sample bias voltage was -2.5\,V and the setpoint current was 0.5\,nA. A swept-frequency sine wave was injected at $U_1$, and the responses were measured at $Y_1$ and $Y_2$. Magnitude and phase of $G(s)$ at each frequency point were obtained by dividing $G_{21}(j\omega)$ over $G_{11}(j\omega)$. The amplitude of swept-frequency sine wave should be kept small to avoid excitation of resonance frequencies and inadvertent changes to the atomic structure of the surface beneath the tip.

\begin{figure}[!tbh]
    \centering
	\includegraphics[width=1\linewidth]{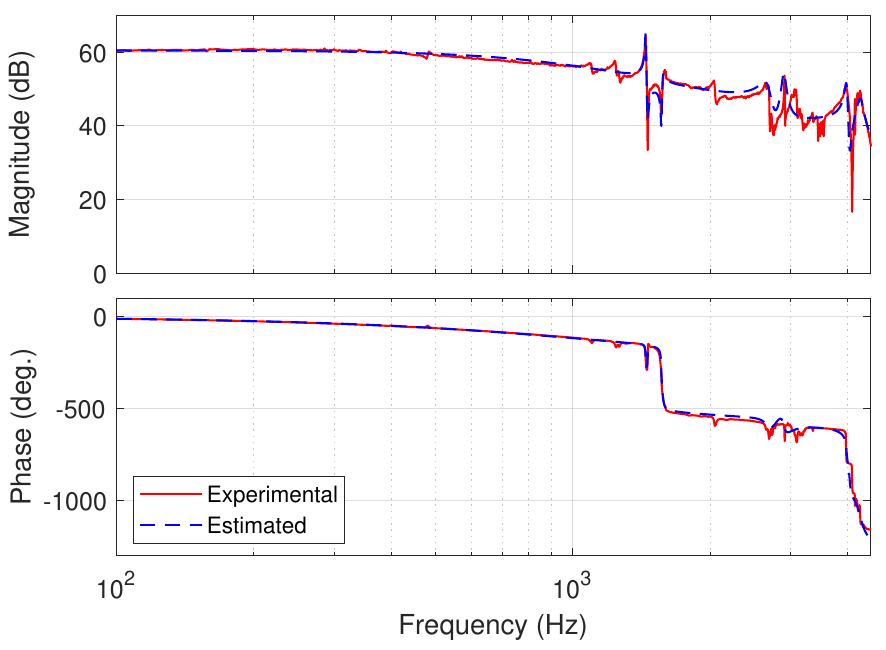}
	\caption{Experimentally obtained frequency response of the system in the constant current imaging mode and the FRF of the identified transfer function.
	\label{fig:FRF_tunn}}
\end{figure}

Figure~\ref{fig:FRF_tunn} shows an experimentally obtained frequency response of the STM and the fitted transfer function. A 14\textsuperscript{th}-order model is fitted to the data up to \mbox{4500\,Hz}. 
\begin{comment}
Poles and zeros of the transfer function
\begin{equation} \label{eq:identified_model}
	G(s) = k_{g}\,G_0(s)=
	    k_{g}\,\frac{(s-z_1)(s-z_2)...(s-z_m)}{(s-p_1)(s-p_2)...(s-p_n)}
\end{equation}

\noindent are listed in Table \ref{table:sysid}. Gain of the transfer function, $k_g$ is $195.2735 \times 10^4$.  
\end{comment}
The root-mean-square error (RMSE) between the identified model and the experimental frequency response is 2.42\,dB.
\begin{comment}
\begin{table}[tbh!]
\centering
\caption{Poles and zeros of identified transfer function for the constant current imaging mode.}
\begin{tabular}{ |c|c|c| }
 \hline
  & Poles & Zeros \\ 
 \hline
 1 & $-477.41 \pm 26802.82i$ & $1617.51 \pm 27543.08i$ \\ 
 2 & $-156.50 \pm 25006.70i$ & $128.58 \pm 25394.91i$ \\  
 3 & $-182.21 \pm 18220.33i$ & $-918.60 \pm 18349.05i$ \\  
 4 & $-493.88 \pm 16778.01i$ & $-484.43 \pm 17401.97i$ \\  
 5 & $-53.61 \pm 9876.76i$ & $20.55 \pm 9824.99i$ \\  
 6 & $-21.17 \pm 9077.12i$ & $-28.51 \pm 9163.22i$ \\  
 7 & $-4951.66 \pm 3208.89i$ & $15584.38$ \\ 
 \hline
\end{tabular}
\label{table:sysid}
\end{table}
\end{comment}

\begin{figure}[!tbh]
    \centering
	\includegraphics[width=1\linewidth]{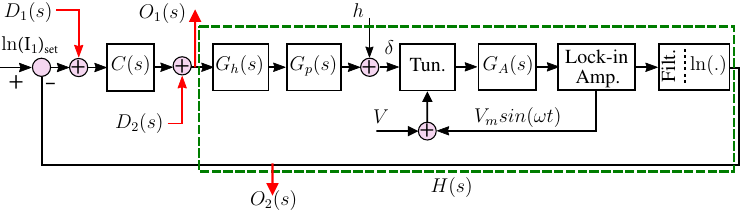}
	\caption{
                Block diagram of the STM z-axis control system operating in constant differential conductance imaging mode. The four underlying transfer functions are $H_{ij}(s)= {O_i(s)}/{D_j(s)}$.
	\label{fig:dIdV_block_diagram_sysid}}
\end{figure}

The block diagram of the  STM's z-axis control system operating in  constant differential conductance imaging mode is depicted in Figure~\ref{fig:dIdV_block_diagram_sysid}. This is rather similar to the constant current imaging mode feedback loop in Figure~\ref{fig:block_diagram_sysid}. However, in this mode, we use a lock-in amplifier to produce a modulation signal that is added to the dc bias voltage of the sample. The resulting current is sent back to the lock-in amplifier and is compared with the modulation signal. The current is demodulated into its in-phase and quadrature components. The feedback loop is then closed on the in-phase component. For a small amplitude modulation signal, the in-phase component is proportional to \Ifd{}.
The open-loop z-axis dynamics of the system in this mode is captured by transfer function $H(s)$. The frequency response of $H(s)$ is measured when the tip is engaged with the surface and the feedback loop is closed. Similar to the constant current case, frequency responses are obtained from the inputs $D_1(s)$ and $D_2(s)$ to the outputs $O_1(s)$ and $O_2(s)$.  $H(j\omega)$ is then obtained as
\begin{equation}
\label{eq:dIdV_closedloop_frf}
	H(j\omega) = \frac{H_{22}(j\omega)}{H_{12}(j\omega)}=
	\frac{H_{21}(j\omega)}{H_{11}(j\omega)}.
\end{equation}

\begin{figure}[!tbh]
    \centering
	\includegraphics[width=1\linewidth]{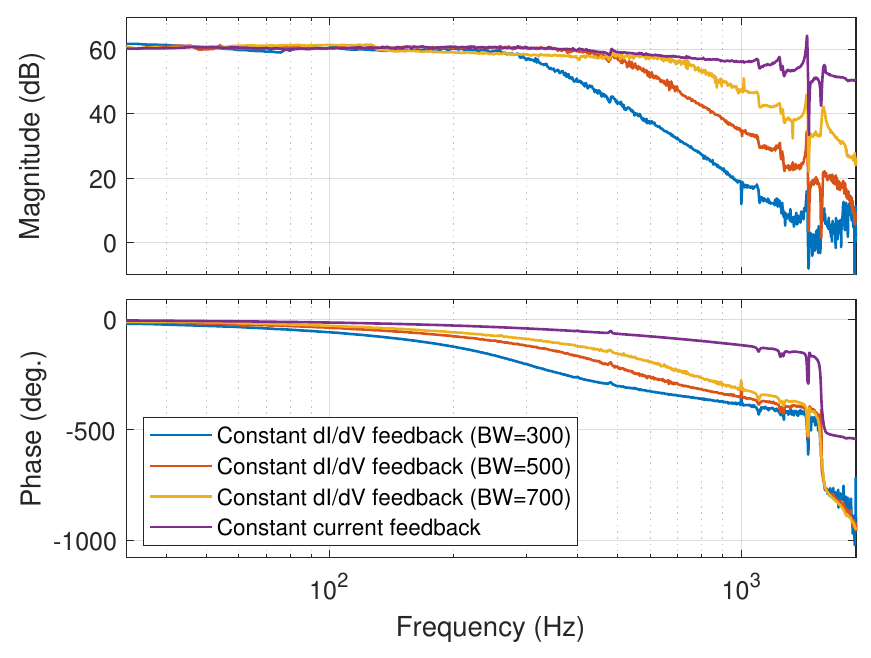}
	\caption{Experimentally obtained frequency responses of the STM system operating in constant differential conductance mode. A \mbox{2\,kHz} modulation voltage with  amplitude of \mbox{0.8\,V} was added to the \mbox{-2.5\,V} dc bias voltage of the sample. The resulting current was demodulated into in-phase and quadrature components at the fundamental frequency. The feedback loop was closed on the in-phase component with the setpoint value of \mbox{0.5\,nA}. The frequency responses are obtained for the lock-in amplifier LPF cutoff frequencies of \mbox{300\,Hz}, \mbox{500\,Hz}, and \mbox{700\,Hz}. FRF of the system is also plotted for comparison.
	\label{fig:dIdV_sysID}}
\end{figure}

To obtain the frequency response of our STM in constant differential conductance mode, a swept-frequency sine wave was injected at $D_1$ and the system responses were measured at $O_1$ and $O_2$, as shown in Figure~\ref{fig:dIdV_block_diagram_sysid}. Magnitude and phase of $H(s)$ at each frequency point were obtained by dividing $H_{21}(j\omega)$ over $H_{11}(j\omega)$. The bandwidth of $H(s)$  depends on the lock-in amplifier properties, particularly on its LPF. Figure~\ref{fig:dIdV_sysID} shows the open-loop frequency responses for three different LPF cutoff frequencies. The open-loop system bandwidth is clearly determined by the LPF cutoff frequency. The LPF rejects out-of-bandwidth noise and disturbances. Hence, lowering the LPF bandwidth enhances the SNR of demodulated signals. This, however, may come at the cost of having to operate the system at a lower scan speed on occasions. Although, since most commercial STM loops operate at bandwidths below 100-200~Hz, this should not be an issue in most cases.

\begin{figure}[!tbh]
    \centering
	\includegraphics[width=1\linewidth]{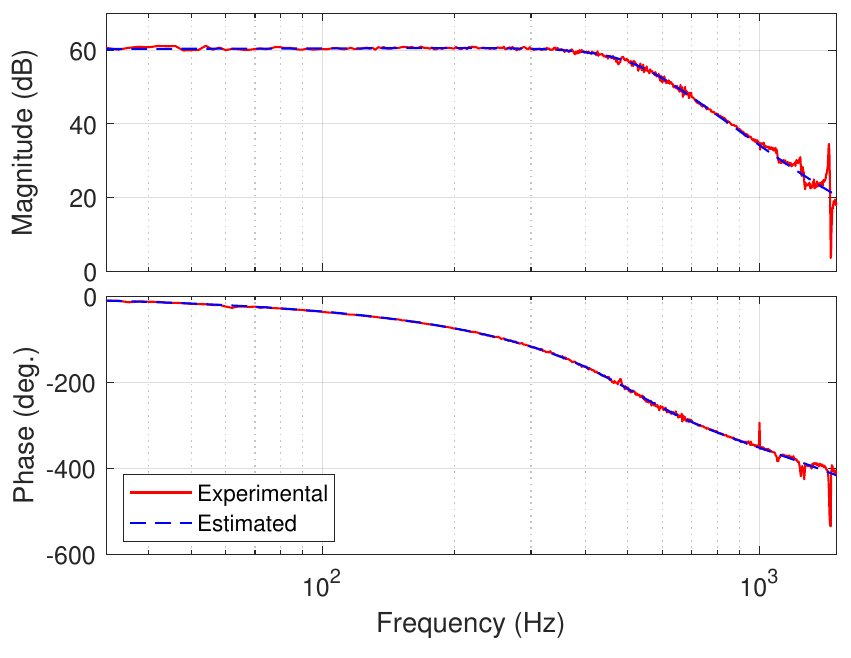}
	\caption{Experimentally obtained frequency response of the system in  constant differential conductance mode plotted together with the FRF of the identified transfer function. A fifth-order transfer model was fitted to the data. The bandwidth of the lock-in amplifier LPF was \mbox{500\,Hz}.  
	\label{fig:dIdV_tfest}}
\end{figure}

A transfer function of the open-loop system is estimated from the frequency response data. Figure~\ref{fig:dIdV_tfest} shows the experimentally obtained frequency response, along with the estimated transfer function. A fourth-order LPF with the cutoff frequency of \mbox{500\,Hz} is used in the lock-in amplifier. A fifth-order model is fitted on the data over \mbox{1900\,Hz}.
\begin{comment}
Poles and zeros of the identified transfer function 
\begin{equation} \label{eq:dIdV_identified_model}
	H(s) = 
	    k_{h}\,\frac{(s-z_1)(s-z_2)...(s-z_m)}{(s-p_1)(s-p_2)...(s-p_n)}
\end{equation}

\noindent are listed in Table~\ref{table:dIdV_sysid}. Gain of the transfer function is $k_h$ of $26666.028$.

\begin{table}[tbh!]
\centering
\caption{Poles and zeros of identified transfer function for the constant \Ifd imaging mode.}
\begin{tabular}{ |c|c|c| }
 \hline
  & Poles & Zeros \\ 
 \hline
 1 & $-1238.46 \pm 2997.79i$ & $5894.26 \pm 8884.52i$ \\ 
 2 & $-2065.07 \pm 1567.28i$ & $24521.14$ \\  
 3 & $-125.86$ & $-120.26$ \\
 \hline
\end{tabular}
\label{table:dIdV_sysid}
\end{table}
\end{comment}

\subsection{Closed-Loop Stability and Performance}
\label{subsubsec:constant_current_stability}
The PI controller is the most frequently used feedback controller in two of the most widely used scanning probe microscopes (SPM): STM, and atomic force microscopes (AFM) (see ``How does an Atomic Force Microscope Work?"). Simple structure, ease of implementation, and its robust performance over a wide range of operating conditions makes the PI controller suitable for SPMs. The controller is defined as
\begin{equation} \label{eq:PI_controller}
	C(s) = k_i\,\left(\frac{1}{s}+\frac{1}{\omega_c}\right)
\end{equation}

\noindent where $k_i$ is the integrator gain and $\omega_c$ is proportional to the corner frequency of the controller. The closed-loop system performance and stability are determined by the PI controller parameters.

The identified open-loop model of the STM can be used to properly design a PI controller. In \cite{Farid_TCST}, the following three criteria are suggested for the STM PI controller design.

\renewcommand{\theenumi}{\roman{enumi}}
\renewcommand{\labelenumi}{\theenumi)}

\begin{figure}[!tbh]
    \centering
	\includegraphics[width=1\linewidth]{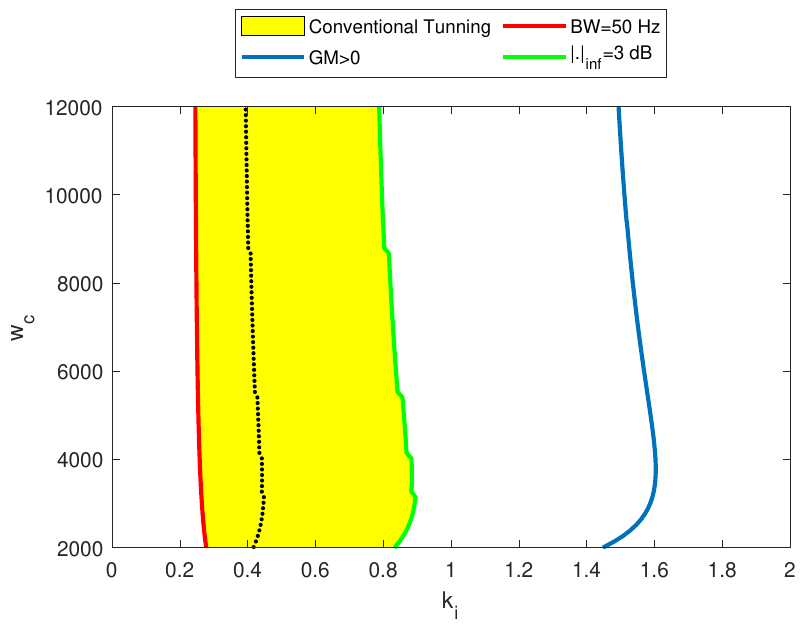}
	\caption{Stable PI controller area for the constant \Ifd{} imaging. To satisfy the required stability and performance requirements, controller parameters should be chosen from the yellow colored area.}
	\label{fig:dIdV_stability}
\end{figure}

\begin{enumerate}
\item Stability margin: The critical integrator gain $k_i$ that brings the closed-loop system to the margin of instability.
\begin{comment}
is determined by finding the gain margin of the following loop transfer function for a given $\omega_c$
\begin{equation} \label{eq:loop_TF}
	G_{lp}(s) = \left(\frac{1}{s}+\frac{1}{\omega_c}\right)\,G(s).
\end{equation}
\end{comment}
\item Imaging bandwidth: In an STM, the probe is scanned over the surface in a raster pattern. This pattern is generated by applying a triangular signal to one axis and a slow ramp signal to the other. As a rule of thumb, the imaging bandwidth should be more than 10 times larger than the scan frequency \cite{Farid_TCST}. The imaging transfer function $G_{img}(s)$ is defined as the transfer function from the topography input $h$ to the controller output.

\item Ringing attenuation: The power spectrum of a triangular signal contains odd harmonics of the fundamental frequency. High-frequency components of the triangular reference signal as well as external disturbances and noise can excite resonances of the piezoelectric tube scanner. These induced vibrations negatively affect imaging and stability performance of the STM. Therefore, limiting the $H_{\infty}$ norm of the imaging transfer function can improve the closed-loop performance.
\end{enumerate}
Based on these criteria, a range of integrator gains $k_i$ can be obtained for each $\omega_c$. For the STM system operating in constant differential conductance imaging mode, this is shown in Figure~\ref{fig:dIdV_stability}. The stability margin criterion determines maximum permissible $k_i$ values marked by the blue curve. The minimum imaging bandwidth is determined by scan parameters. Higher scan speeds necessitate a higher imaging bandwidth. Similarly, a sample with small features requires a higher number of pixels within each nanometer of image and higher imaging bandwidth. In our case, we defined the minimum imaging bandwidth of \mbox{50\,Hz}, which imposes a lower bound on the $k_i$, shown with the red curve. Furthermore, the $H_{\infty}$ norm requirement of  \mbox{3\,dB} restricts the admissible range of  integrator gain as shown by the green curve. To satisfy the above three criteria, the PI controller parameters should be chosen to be in the highlighted yellow area in Figure~\ref{fig:dIdV_stability}. The black dashed line represents the recommended integrator gain, which is half of the upper bound. The aforementioned design criteria provide adequate gain and phase margins while ensuring that the first resonance frequency of the closed-loop system is attenuated so that inadvertent ringing would not adversely affect the imaging performance. These criteria can be employed to design PI controllers for both constant current and constant differential conductance imaging modes.
%%%%%%%%%%%%%%%%%%%%%%%%%%%%%%%%%%%%%%%%%%%%%%%%%

%%%%%%%%%%% Sec05 %%%%%%%%%%%%%%%%%%%%%%%%%%%%%%%
\section{Self-tuning PI Controller}
\label{sec:LBH_estimation_PI_tuning}
The dc gain of the open-loop transfer function $G(s)$ in Figure~\ref{fig:block_diagram_sysid} is 
\begin{equation} \label{eq:dc_gain}
	\rVert G(0) \rVert=
	k_g \rVert G_0(0) \rVert=
	1.025 \sqrt{\phi}\,k_{hva}\,k_{piezo}
\end{equation}

\noindent where $k_{hva}$ and $k_{piezo}$ are the dc gains of the high voltage amplifier and the piezoelectric tube scanner, respectively. These two parameters are constant; however, the LBH is a strong function of local electronic charges on the surface.

It was first noticed in \cite{Farid_RSI} that variations in the LBH proportionally change the dc gain of the STM  system as the tip scans over the surface. Equation (\ref{eq:dc_gain}) clearly shows this. These LBH variations are large enough to destabilize the closed-loop system \cite{Farid_TCST}. As a remedy, an online adaptive algorithm was proposed in \cite{Moheimani_2019_LBHpatent} to adjust the PI controller gain based on the LBH variations. In this section, we discuss two different methods to estimate LBH variations in real-time and a self-tuning PI controller to compensate for the resulting variations in the dc gain.

\subsection{Local Barrier Height Estimation}
\label{subsec:LBH_estimation}
LBH provides important information about  local electronic charges on the surface. As was shown in \eqref{eq:barrier_height}, LBH is proportional to the $\dv*{(\ln{I})}{\delta}$. Gap modulation method (also known as $\dv*{I}{z}$ spectroscopy) is commonly used to estimate LBH variations \cite{Lang_1988_GapModulation, Binnig_2000_GapModulation, Maeda_2004_GapModulation}. In this method, a high-frequency dither voltage with a fixed amplitude is added to the the piezoelectric tube actuator voltage, resulting in small tip oscillations ($\Delta z$) that are assumed to be of a constant amplitude. Consequently, oscillations with the same frequency appear on the logarithm of current ($\Delta (\ln{I})$). A demodulator, commonly a lock-in amplifier, is then used to estimate the amplitude of induced oscillations from which $\dv*{(\ln{I})}{z}$ is calculated. In this approach, it is assumed that the resulting high-frequency components are completely filtered out by the controller from the feedback signal. However, these signals could find their way to the z-axis actuation signal, particularly when the PI controller gains are high or the dither voltage has a relatively large amplitude. As a result, the tip-sample gap  is affected by these oscillations violating the initial assumption of constant tip oscillations, for example, a 0.1\,\AA{ngstrom} change in tip-sample distance results in a 20\% change in tunneling current \cite{Book_Voigtlaender}. Thus, the estimated LBH obtained with this method tends to be inaccurate \cite{Farid_TCST}.

A modified LBH estimation method was proposed in \cite{Farid_TCST} that works based on estimating the STM open-loop dc gain. This is very similar to the open-loop frequency response measurement technique that was previously described in detail. A high-frequency modulation signal is added to the closed-loop system at the setpoint $U_1(j\omega)$, as shown  in Figure~\ref{fig:block_diagram_sysid}. Then, the STM response at the input of the high-voltage amplifier $Y_1(j\omega)$ and at the output of the preamplifier $Y_2(j\omega)$ are
\begin{align}
	Y_{1}(j\omega) &= \frac{C(j\omega)}
        {1+C(j\omega)G(j\omega)}\,U_{1}(j\omega)
        \label{eq:LBH_Y1U1} \\
	Y_{2}(j\omega) &= \frac{C(j\omega)G(j\omega)}
	    {1+C(j\omega)G(j\omega)}\,U_{1}(j\omega)
	    \label{eq:LBH_Y2U1}.
\end{align}

\noindent The STM  response at the modulation frequency can be obtained by dividing $Y_{2}(j\omega)$ by $Y_{1}(j\omega)$ as
\begin{equation} \label{eq:response_at_jw}
	\frac{Y_{2}(j\omega)}{Y_{1}(j\omega)} = G(j\omega).
\end{equation}

\noindent Therefore, the magnitude of ${\rVert Y_{2}(j\omega) \rVert}/{\rVert Y_{1}(j\omega) \rVert}$ is proportional to the LBH, that is,
\begin{equation} \label{eq:LBH}
	\frac{\rVert Y_{2}(j\omega) \rVert}{\rVert Y_{1}(j\omega) \rVert}=
	\rVert G(j\omega) \rVert \propto
	\rVert G(0) \rVert \propto
	\sqrt{\phi}.
\end{equation}

\noindent Alternatively, the modulation signal could be added to the controller output.

In \cite{Farid_TCST}, two lock-in amplifiers were employed to estimate $\rVert Y_{1}(j\omega) \rVert$ and $\rVert Y_{2}(j\omega) \rVert$. This provides a more accurate LBH estimation, since the effect of the feedback signal on the $\Delta z$ is also considered. LBH is less correlated with the topography in this modified LBH estimation method. Frequency of the modulation signal should not be very high to achieve a good SNR in measuring $\rVert Y_{2}(j\omega) \rVert$. Otherwise, the high-frequency roll-off of the closed loop response could lead to noisy measurements. Furthermore, this frequency should be selected so that it is not in the proximity of piezotube's resonances. Moreover, the modulation frequency should not be lower than the closed-loop imaging bandwidth to prevent disturbing the topography signal. The imaging bandwidth of conventional STM is typically in the range of tens of hertz to a few hundred hertz.
\begin{comment}
\begin{figure}[!htb]
	\noindent
	\centering
	\includegraphics[width=1.0\linewidth]{LBH_Farid_modified.jpg}
	\caption{LBH images obtained using the modified (middle) and conventional (bottom) LBH estimation methods. Experiments were performed with fixed PI gains. The profile B on the middle image (modified LBH estimation method) shows LBH variations of 21.6\% due to the topography effects. The same profile line (profile C on the bottom image) is drawn on the LBH image that was obtained with the conventional method. In this case the LBH changes due to the topography features is 60.4\%. This experiment shows that LBH estimations with the modified LBH estimation method is less correlated to the topography. Figure reprinted with permission from \cite{Farid_RSI}, copyright 2018, AIP.}
	\label{fig:LBH_Imaging}
\end{figure}  
\end{comment}

\subsection{PI Controller Gain Adaptation}
\label{subsec:PI_Tuning}
We have already shown that the dc gain of an STM changes during a scan due to the LBH variations. Also, two different methods were presented for real-time estimation of the LBH. Large LBH variations can cause significant changes in the parameter $k_g$ in \eqref{eq:dc_gain}. This may result in closed-loop  instabilities causing a tip-sample crash, which is common in STM. The closed-loop system gain can be kept constant by adapting the controller gain based on the estimated LBH variations according to
\begin{equation} \label{eq:PI_adapting}
	(k_i)_{new}=k_i\frac{(\widetilde{k}_g)_{des}}{(\widetilde{k}_g)_{est}}
\end{equation}

\noindent where $(\widetilde{k}_g)_{est}$ and $(\widetilde{k}_g)_{des}$ are the estimated and the desired dc gains of the open-loop system, respectively \cite{Farid_TCST, Farid_RSI}. Parameter $(\widetilde{k}_g)_{des}$ is determined by averaging $\dv*{(\ln{I})}{z}$ signal prior to commencing a scan. Parameter $(\widetilde{k}_g)_{est}$ is the estimated value of $\dv*{(\ln{I})}{z}$ during the scan.

The LBH can change from one atom to another during a scan. Therefore, it needs to be estimated in real time, and the controller gains should be constantly adjusted based on the estimated value of the LBH. Bandwidth of the LBH estimation algorithm is always slower than the actual physical changes in the LBH. The scanning speed of the STM is slow enough so that multiple data points would be collected over each atom during a scan. This means that the slower speed of the LBH estimation would not be an issue. The order and bandwidth of the LPF integrated into the lock-in amplifier determines the LBH estimation speed and performance. A high-bandwidth LPF enables faster detection of the LBH variations; however, it comes at the cost of a noisier estimation. The closed-loop imaging bandwidth should be smaller than the LBH estimation algorithm bandwidth to ensure that the PI gains can be adjusted fast enough in response to the dc gain variations.
%%%%%%%%%%%%%%%%%%%%%%%%%%%%%%%%%%%%%%%%%%%%%%%%%

%%%%%%%%%%% Sec06 %%%%%%%%%%%%%%%%%%%%%%%%%%%%%%%
\section{Scanning Tunneling Spectroscopy}
\label{sec:scanning_tunneling_spectroscopy}
An important feature of an STM is its ability to obtain energy-resolved atomic-resolution spectroscopic data from a surface. The ultimate goal of an STS is to measure the density of states (DOS) of the surface. This can be achieved by measuring current-voltage (\mbox{I-V}) characteristic of the tunneling junction. A small increase in the voltage $dV$ results in an additional current contribution $dI$ to the total current. This additional current per voltage increase is quantified by \Ifd{}. By assuming that the DOS of the tip $\rho_{t}$ and the transmission factor $T$ are voltage independent in \eqref{eq:SimplifiedTunn}, the differential conductance \Ifd{} is obtained as
\begin{equation}\label{eq:dIdV}
	\dv{I}{V} = R\,\dfrac{4\pi e^2}{\hbar} \rho_{t}(0) \rho_{s}(ev) T(ev, V, \delta).
\end{equation}

\noindent Therefore, the differential conductance, \Ifd{} is proportional to the density of states of the sample
\begin{equation}\label{eq:LDOS}
	\dv{I}{V} \propto \rho_{s}(ev).
\end{equation}

Various STS methods have been proposed to acquire information on the local electronic properties of the sample \cite{STS_Feenstra}, and many have been integrated into the existing commercial STM systems. The simplest STS method is based on successive scanning of a surface at different positive sample bias voltages \cite{STS_Owen, BiasDependentSTS_01, BiasDependentSTS_02}. This provides a complete map of the surface with information on distinct sample states at different energy levels. Another method known as single-point spectroscopy provides local electronic properties at a specific location on the surface. In this method, the tip is maintained at a fixed height over the desired atom or molecule. Then, the bias voltage is changed with a slow rate and the corresponding tunneling current is recorded. The constructed \mbox{I-V} curve contains important local information  regarding the surface \cite{DB_Wolkow01}, the slope of which provides the \Ifd{} over the desired range of voltages.

Majority of electrons tunnel elastically through the tip-sample junction. However, under certain conditions, a small fraction of electrons can tunnel inelastically. Tunneling electrons can excite an adsorbate vibrational mode when their energy is equal to the excitation energy of the adsorbate. This opens an inelastic channel in parallel with the elastic one  \cite{IETS_01, IETS_02, IETS_03}. This process results in observation of a pair of steps in the \Ifd{} image as well as a peak-and-dip pair in the \Isd{} image at the bias voltage corresponding to the energy of the vibrational mode. STM inelastic electron tunneling spectroscopy (STM-IETS) employs this unique property as a fingerprint to identify the molecular species on the surface. In this spectroscopy method, the \Isd{} signal is produced and plotted along with the XY position of the tip. The lock-in technique can be employed to produce and measure a \Ind{} signal in the STM. This method is introduced in the following section.

\subsection{Experimental realization of STS}
\label{subsec:dIdV_spectroscopy}
\begin{figure}[!htb]
	\noindent
	\includegraphics[width=1\linewidth]{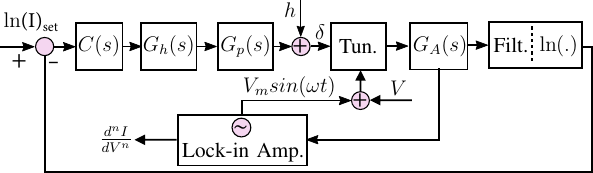}
	\caption{Lock-in method to obtain \Ind. In this method, a modulation voltage is added to the dc bias voltage of the sample. The measured current is then sent to the lock-in amplifier and demodulated into the \Ind.}
	\label{fig:BlockDiag}
\end{figure}

The lock-in technique can provide high-quality \Ind{} STS images of a surface along with the topography image, as shown schematically in Figure~\ref{fig:BlockDiag}. The \Ind{} signal obtained with this method contains considerably less noise compared with the direct numerical differentiation of the \mbox{I-V} curve data. A sinusoidal high-frequency modulation voltage, $V_m\,sin(\omega t)$, is added to the dc bias voltage of the sample $V$, resulting in a capacitive current that leads the modulation voltage by 90\degree{}, in addition to a tunneling current that is a function of the applied  voltage. That is,
\begin{equation}  \label{eq:TotalCurrent}
    \begin{aligned}
	    I_{total} &= I_{cap}+I\\
    	&=CV_{m}\omega\,cos(\omega t)+f(V+V_m\,sin(\omega t)).
    \end{aligned}
\end{equation}

\noindent The Taylor series expansion of $I$ around the bias voltage $V$ is
\begin{equation} \label{eq:Taylor}
	\begin{aligned}
		I&=f(V+V_m\,sin(\omega t))=\sum_{k=0}^{\infty} \frac{V_m^k}{k\,!} \dv[k]{f}{V}  sin^{k}(\omega\,t)\\
		&=1\left(f(V)+\frac{V_m^2}{4} \dv[2]{f}{V} +\frac{V_m^4}{64} \dv[4]{f}{V} +\dots\right)\\
		&+V_m\,sin(\omega t)\left(\dv{f}{V}+\frac{V_m^3}{8} \dv[3]{f}{V} +\frac{V_m^5}{192} \dv[5]{f}{V} +\dots\right)\\
		&-\frac{V_{m}^2}{4}\,cos(2\omega t)\left(\dv[2]{f}{V}+\frac{V_m^4}{12} \dv[4]{f}{V} +\dots\right)\\
		&-\frac{V_{m}^3}{24}\,sin(3\omega t)\left(\dv[3]{f}{V} +\frac{V_m^5}{16} \dv[5]{f}{V} +\dots\right)\\
		&+\frac{V_{m}^4}{192}\,cos(4\omega t)\left(\dv[4]{f}{V}+\dots\right).\\
	\end{aligned}
\end{equation}

\noindent Equation~\eqref{eq:Taylor} can be rewritten  as 
\begin{equation} \label{eq:Taylor2}
	I=I_0+\sum_{n=1}^{\infty}\left(I_{2n-1}sin((2n-1)\omega\,t)+I_{2n}cos(2n\omega\,t)\right).\\
\end{equation}

The lock-in amplifier measures the total current and compares it with the reference modulation signal. Then, it demodulates the current into in-phase and quadrature components at the fundamental frequency of $\omega$ and its harmonics. The in-phase component of current with the reference signal at the fundamental frequency is equal to $I_1$, and the quadrature component at the second harmonic is equal to $I_2$. For a small dither voltage, $I_n$ can be approximated by the lowest-order derivative, that is,
\begin{equation} \label{eq:SmallDitherTaylor}
	I_n\propto \dv[n]{I}{V}.\\
\end{equation}

\subsection{High signal-to-noise ratio spectroscopy}
\label{subsec:MultimodeSTM}
In conventional \Ind{} measurement techniques, the high-frequency modulation voltage introduces high-frequency noise on the current signal that is then projected back to the controller through the feedback loop. This noise disturbs the controller output and negatively affects the resulting topography image. In addition, it can excite the STM resonant frequencies, which may ultimately lead to a tip crash due to high-frequency tip oscillations. To mitigate the adverse effects of this noise on the feedback loop, the amplitude of modulation voltage $V_m$ is constrained to small values. However, this comes at the price of \Ind{} measurements being of a rather low SNR. In practice, it is very difficult to measure anything beyond the second derivative with conventional STS.   

Incorporating notch filters in the STM feedback loop mitigates the negative impact of the induced high-frequency noise due to the addition of a dither signal to the bias voltage \cite{Alemansour_fastSTS}. The center frequency of notch filters should be set to the dither frequency and the first few higher harmonics. Having more notch filters results in better noise attenuation of the tunneling current; however, it comes at the expense of increased implementation complexity. The block diagram of the modified closed-loop STM control system is shown in Figure~\ref{fig:BlockDiag}. The lock-in amplifier generates a modulation voltage that is added to the dc bias voltage of the sample. The resulting current is then sent to the lock-in amplifier to obtain the \Ind{} signal. Unlike the conventional approach, where the logarithm of unfiltered current is fed back to the controller, the current passes through a bunch of notch filters (tuned to $\omega$ and its harmonics, shown as the "Filt" block) before taking the logarithm of current. These filters eliminate the induced high-frequency noise in the current signal.  This modification in the feedback loop makes it possible to apply a higher amplitude modulation signal enhancing the SNR of \Ind{} measurements, while the STM control system continues to function in conventional constant current imaging mode \cite{Hamed_patent, Hamed_STM_03}.
\begin{figure*}
    \centering
    \includegraphics[width=1\textwidth]{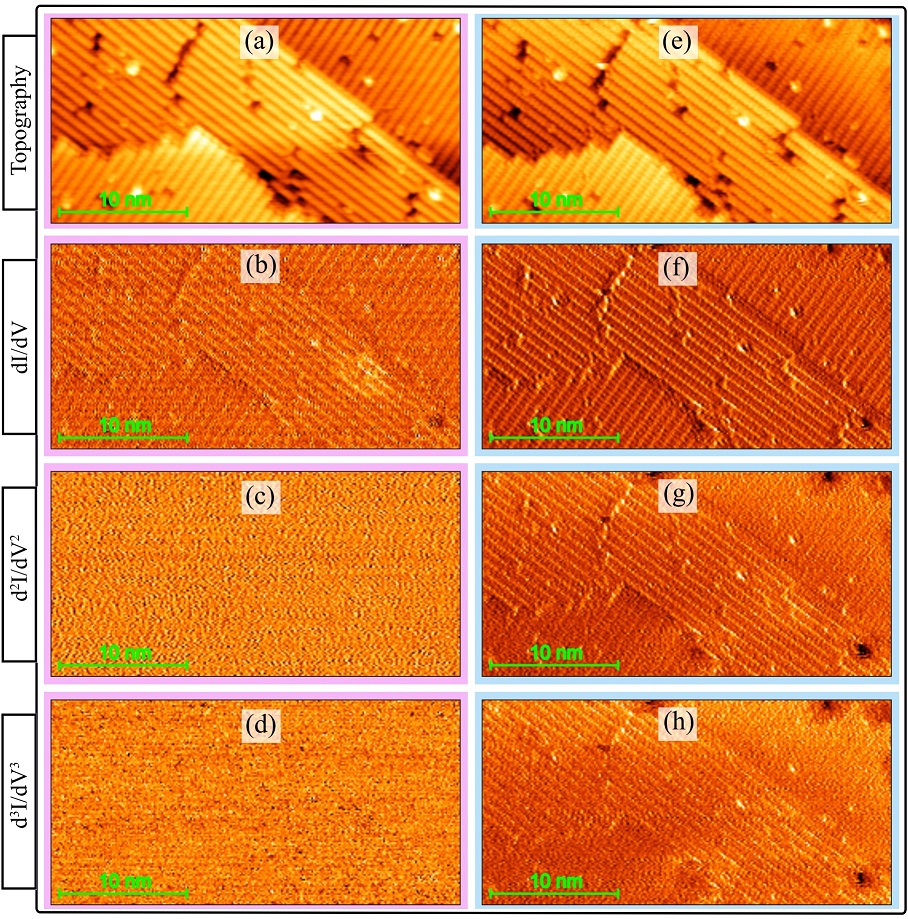}
    \caption{High SNR \Ind{} imaging of a H-passivated silicon surface. A \mbox{2\,kHz} modulation voltage was added to the sample’s \mbox{-2.5\,V} dc bias voltage. (a)-(d) The amplitude of modulation voltage was \mbox{0.1\,V}. In this conventional mode, notch filters were not included in the feedback loop. (e)-(h) The modulation amplitude was increased to \mbox{0.8\,V}, and notch filters were integrated into the feedback loop. Figure reprinted with permission from \cite{Alemansour_fastSTS}, copyright 2021, AIP.}
    \label{fig:Vibrational_STS}
\end{figure*}

Figures~\ref{fig:Vibrational_STS}~(a-d) depict the topography, \Ifd{}, \Isd{}, and \Itd{} images obtained by applying a small amplitude modulation voltage with no notch filters being in the feedback loop, that is, the conventional approach. The quality of \Ifd{} image is quite low, and we can barely see any features in \Isd{} and \Itd{} images. After incorporating notch filters  in the feedback loop and scanning the same area with a higher amplitude modulation signal, we obtain the results plotted in \mbox{Figures~\ref{fig:Vibrational_STS}~(f-h)}. Clearly, these \Ind{} measurements are superior to those obtained by the conventional method.

\subsection{Ultrafast current-voltage spectroscopy}
\label{subsec:fast_CITS}
Current-voltage spectroscopy provides important information on electronic properties of the surface. A single-point \mbox{I-V} curve can be  obtained by freezing the tip height (that is by opening the z-axis feedback loop) and slowly sweeping the bias voltage over the desired range. The tunneling current is then measured and plotted versus the bias voltage \cite{STS_IndividualAtoms}. The differential conductance can also be obtained through numerical differentiation of the \mbox{I-V} curve data. A complete map of the surface can be produced by repeating this process for every pixel of the image. This method is called current-imaging tunneling spectroscopy (CITS). Although conceptually this method seems quite simple, in practice, it requires a very stable STM system with a low lateral drift that can handle the long dwelling times at each pixel \cite{bonnell}. Therefore, CITS is a very slow method, requiring several hours of operation to obtain the I-V measurements over  a small area of the surface. Furthermore, this conventional approach to CITS is known to fail repeatedly, since freezing the tip height by disengaging the feedback controller leaves the tip unprotected against noise, vibration, and drift.

An ultra-fast spectroscopy technique was introduced in \cite{Alemansour_fastSTS} to address the above issues with CITS. This method makes it possible to acquire an \mbox{I-V} curve for every pixel of an image at a remarkably fast rate. A lock-in amplifier generates a high-frequency sine signal that is  added to the sample bias voltage. Due to the high-frequency nature of the modulation signal, a significant part of the total current (that is the preamplifier output signal) is capacitive. The current that is used to construct an \mbox{I-V} curve should only contain tunneling components. Therefore, the capacitive current should be taken out of the total current. We showed in \eqref{eq:TotalCurrent} that the resulting capacitive current has an amplitude of $C V_m \omega$ and leads the tunneling current by $90^\circ$.  Hence, its amplitude can be easily measured by a lock-in amplifier. The preamplifier output is sent to the lock-in amplifier and is demodulated into  in-phase and the quadrature components at the fundamental frequency. The quadrature component is the amplitude of the capacitive current, that is, $C V_m \omega$. The tunneling current is then calculated by subtracting this from the total current. This measured signal is plotted against the modulation voltage amplitude to construct \mbox{I-V} curves during a scan. A current image of the surface for a selected tip-sample bias voltage can be constructed from the obtained \mbox{I-V} curves data. This is realized by extracting the tunneling current value corresponding to the selected voltage for each of the \mbox{I-V} curves and then plotting them against  the lateral position of the tip.

The bias voltage can even be zero in the ultra-fast spectroscopy method \cite{Alemansour_fastSTS}. Here, every step is identical to the case where the bias voltage is not zero, with the only difference being that the feedback loop is now closed on the in-phase component, that is, $I_1$ in \eqref{eq:Taylor2}. The logarithm of this signal is compared with the setpoint and the error signal is applied to the controller. The output of the controller  represents the surface topography. The zero bias voltage makes it possible to have a symmetric voltage about the origin for the \mbox{I-V} curves. As a result, a modulation signal with a smaller amplitude can be applied to the sample. This is beneficial when we work with samples that can potentially change with a high bias voltage. For example, suppose that the voltage range of our \mbox{I-V} curves needs to be \mbox{-2.5\,V} to \mbox{+2.5\,V}. For a zero-volt dc bias voltage, the modulation amplitude should be \mbox{2.5\,V} to sweep the desired voltage range. However, for a \mbox{-2.5\,V} dc bias voltage, the modulation amplitude must increase to \mbox{5\,V}, that is, the total tip-sample bias voltage of \mbox{-7.5\,V} to \mbox{2.5\,V}. For a H-terminated silicon surface, this considerably increases the risk of inadvertent hydrogen depassivation.

\begin{figure}[tbh]
    \centering
	\includegraphics[width=1\linewidth]{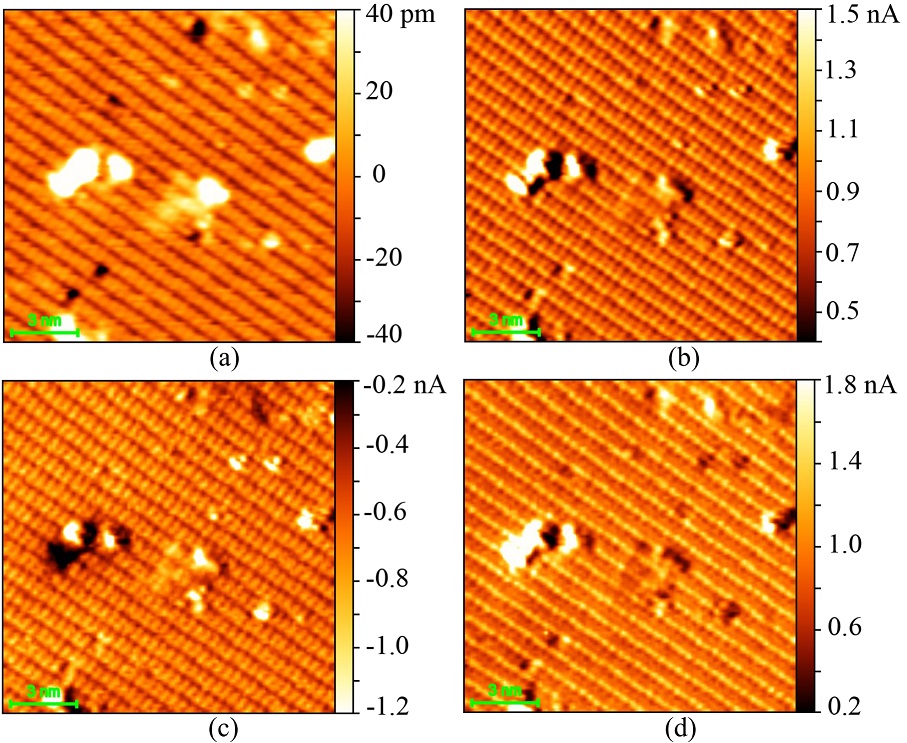}
	\caption{STM and STS images of an H-passivated silicon surface. The lock-in amplifier applies a modulation signal with the amplitude of \mbox{2.5\,V} and the frequency of \mbox{2\,kHz} to the sample. The feedback loop was closed on the in-phase component of the measured current with the modulation signal at the fundamental frequency. All of the images were obtained simultaneously with the tip speed of \mbox{100\,nm/s}. (a) The topography image. (b) The feedback signal. (c) The tunneling current image at the voltage of \mbox{-\,2.11\,V}. (d) The tunneling current image at the voltage of \mbox{+\,2.26\,V}.
	\label{fig:Fast_IV}}
\end{figure}

Figure~\ref{fig:Fast_IV} depicts the STM and STS images of a Si(100)-2\texttimes{}1:H passivated surface that were obtained with this method. In this experiment, our method was programmed into a ZyVector STM Control System  \cite{ZyvexWeb}.  Every task was automated within the software, including the digital lock-in amplifier that generates a reference signal with the amplitude and frequency of \mbox{+2.5\,V} and \mbox{2\,kHz}. This signal is added to the sample with zero dc bias voltage. The resulting current is amplified by a FEMTO LCA-400K-10M transimpedance amplifier, which is then sent to the Zyvector. The lock-in amplifier demodulates the signal into in-phase and  quadrature components at the fundamental frequency. The feedback loop is closed on the natural logarithm of the in-phase component to keep $I_1$ at \mbox{1\,nA}. At each sampling time, the capacitive component is subtracted from the total current, the result of which is recorded as the tunneling current. It is then plotted against the bias voltage to construct \mbox{I-V} curves during a scan. In this experiment, a \mbox{15\,nm $\times$ 15\,nm} area is scanned at a tip speed of \mbox{100\,nm/s}. Each image in Figure~\ref{fig:Fast_IV} has a pixel size of \mbox{120 $\times$ 120}. The controller output represents the topography of the surface and is shown in Figure~\ref{fig:Fast_IV}\,(a). The image of $I_1$ signal is plotted in Figure~\ref{fig:Fast_IV}\,(b). Images of current slices at bias voltages of \mbox{-2.11\,V} and \mbox{+2.26\,V} are shown in Figures~\ref{fig:Fast_IV}\,(c) and \ref{fig:Fast_IV}\,(d), respectively. With this method, an spectroscopic map of the surface can be constructed at least 1500 times faster than the conventional CITS method \cite{CITS_time, Alemansour_fastSTS}.

%%%%%%%%%%%%%%%%%%%%%%%%%%%%%%%%%%%%%%%%%%%%%%%%%

%%%%%%%%%%% Sec07 %%%%%%%%%%%%%%%%%%%%%%%%%%%%%%%
\section{Hydrogen Depassivation Lithography}
STM-based hydrogen depassivation lithography (HDL) has been the subject of intensive research in  recent years \cite{Avouris_CPL_1996_HDL,Hersam_2000_FCL,Lyding_1994,Lyding_Abeln_1994,Avouris_1996_HDL,Dagata_1990_HDL}. In this technology, an STM tip  injects current into the Si-H bond on a hydrogen-passivated silicon surface to selectively remove H atoms from the surface by breaking the chemical bonds. This results in the formation of a silicon dangling bond (DB) wherever an H atom is removed. The exposed dangling bond is highly reactive and quickly forms a covalent bond with other species of atoms or molecules when they come into contact with the surface. This provides an opportunity to place dopant atoms in silicon, with atomic precision, to fabricate innovative silicon quantum electronic devices \cite{Ruess_2004_doping,Buch_2015_doping,RevModPhys.85.961} (see ``Atomically Precise Manufacturing for Si Quantum Computing").

\subsection{Conventional HDL Modes}
HDL is commonly performed with a positive sample bias voltage. There are two distinct HDL modes: atomic precision \cite{Soukiassian_2003_APL, Randall_2019_HDL} and field emission \cite{Shen_1995_FEM, Avouris_1996_HDL}. In atomic precision (AP) mode, the dc bias voltage is less than \mbox{6\,V} and single hydrogen atoms can be removed from the surface selectively.  A higher bias voltage is required to perform lithography in  field emission (FE) mode. The FE mode enables high throughput lithography. However, the resolution is typically limited to  \mbox{5\,nm}. HDL can also be performed at a negative sample bias voltage \cite{Stokbro_1998_HDL}. However, this would require a larger bias voltage and setpoint current increasing the chance of a tip-sample crash.

To perform lithography, bias voltage and  setpoint current are switched from  imaging to  lithography values. This changes the tip-sample distance and, hence, the barrier height. We explained earlier in \eqref{eq:dc_gain} that a change in barrier height will change the open-loop gain of the STM system. We showed in \cite{Hamed_STM_02} that this change in system gain may be large enough to destabilize the closed-loop system. A similar phenomenon occurs during lithography when an H atom leaves the surface exposing a Si dangling bond under the tip. The instantaneous transition from tunneling into an H atom to a Si atom results in a significant step change in system gain that could destabilize the closed-loop system \cite{Farid_RSI, Farid_TCST}. LBH estimation-based tuning of the STM controller, explained above, can address this issue. 

\subsection{Constant Differential Conductance mode STM-based HDL}
\begin{figure}[htb]
	\noindent
	\centering
	\includegraphics[width=1\linewidth]{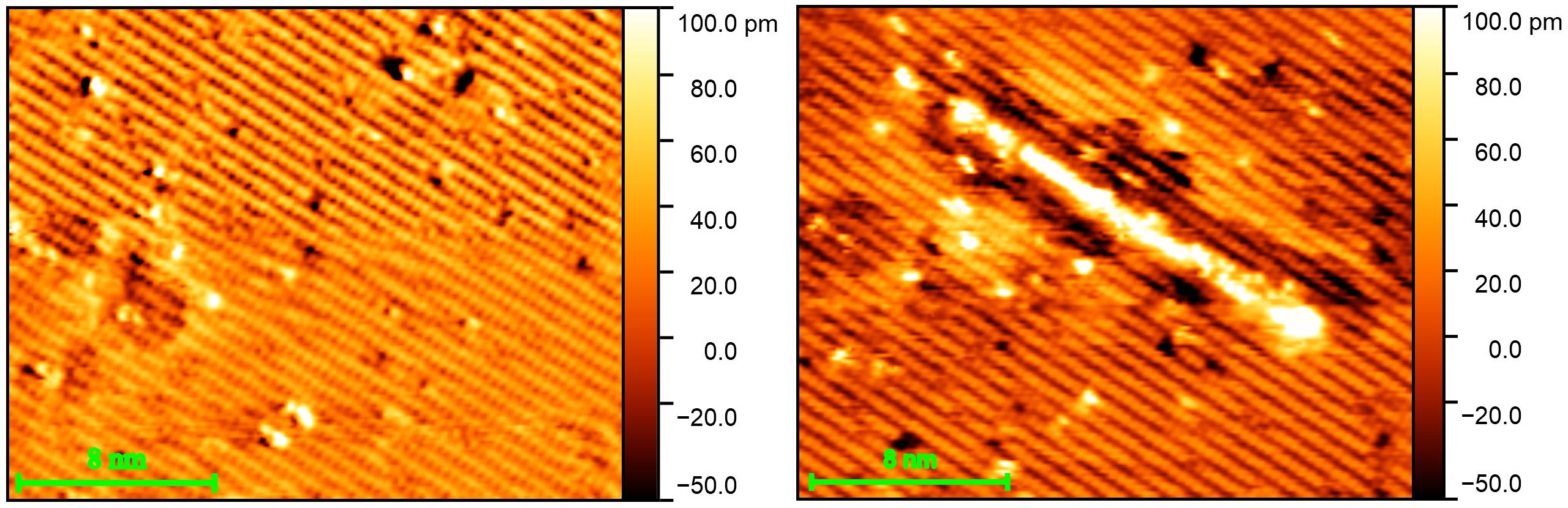}
	\caption{Lithography performed along a dimmer row in  constant \Ifd{} imaging mode. Left image shows the surface before the lithography and right image, immediately after.}
	\label{fig:dIdV_lithography}
\end{figure}

Hydrogen depassivation lithography can be performed while the feedback loop is closed on the \Ifd{} signal. Figure~\ref{fig:dIdV_lithography} shows HDL performed along a dimmer row with \Ifd{} feedback. In this experiment, amplitude of the modulation voltage was \mbox{0.8\,V} and modulation frequency was \mbox{2\,kHz}. The feedback loop was closed on the \Ifd{} signal for both imaging and lithography. The imaging was performed with the \Ifd{} setpoint of \mbox{0.9375\,nA/V}, the tip speed of \mbox{100\,nm/sec}, and the bias voltage of \mbox{-2.5\,V}. For lithography, the setpoint and the bias voltage were switched to \mbox{2.5\,nA/V} and \mbox{+3\,V}, respectively, and the tip was moved along the dimmer row with the speed of \mbox{5\,nm/sec}.

\subsection{Feedback-Controlled Lithography}
 Error-free formation of dangling bond structures on a H-passivated Si surface is a critical step in the fabrication of solid-state quantum devices with atomic precision \cite{achal_2018_lithography}. The conventional approach to HDL often fails to produce the requisite patterns with 100\% precision. Either the method fails to remove all hydrogen atoms selected for desorbtion, or it reults in the formation of random dangling bonds in the vicinity of the feature. These seemingly small errors in the lithography can lead to a complete failure of the device fabrication process.

Lyding proposed a method known as feedback-controlled lithography (FCL) to improve the reliability of the AP lithography process \cite{Hersam_2000_FCL, Basu_2004_FCL}. As soon as an H atoms breaks bond with Si and leaves the surface, the local barrier height (and thus, the flow of tunneled electrons from the resulting dangling bond to the tip undergo a step change).  This results in a sudden increase in the tunneling current, since the LDOS of the Si atom is higher than  hydrogen. In response to this change in measured current, the PI controller moves the tip further away from the surface to maintain the setpoint current. The FCL method uses this desorption signature to detect the moment an Si-H bond is broken and a dangling bond is formed \cite{Moller_2017_FCL}. In this method, either the control signal or the tunneling current is actively monitored, and the lithography process is terminated as soon as a desorption event is detected. This procedure  reduces the likeliness of unintentional depassivation of nearby H atoms, and thus random formation of dangling bonds. 

\subsection{Voltage-Modulated Feedback-Controlled Lithography}
As we previously stated, switching between imaging and lithography modes alters the dc gain of the open-loop system adversely affecting the closed-loop stability of the system. We introduced the voltage-modulated lithography technique in \cite{Hamed_STM_01, Hamed_STM_02} to mitigate this problem. In this method, there is no need to change the bias voltage or the setpoint current value to transition from imaging to lithography mode. Instead, the lithography is initiated by adding a high-frequency sinusoidal voltage to the dc bias voltage of the sample. This provides the required energy to remove a hydrogen atom while the STM continues to function in the imaging mode. The elimination of switching step in this method reduces the likelihood of a tip-sample crash. The high-frequency modulation signal induces high-frequency components on the current signal at the integer multiples of the modulation frequency. Therefore, it is important to select a modulation frequency that is higher than the imaging bandwidth to minimally disturb the topography signal. In addition, the modulation frequency should be selected such that neither it nor its harmonics excite resonant dynamics of the piezoelectric tube scanner. Since the amplitude of the modulation signal could become relatively high, the amplified current should be passed through a set of notch filters to attenuate any excessive high-frequency components on the current signal. 

The precision of this lithography method has been further improved by combining it with  Lyding's FCL method in \cite{Hamed_STM_02}. This novel HDL technique is called voltage-modulated feedback-controlled lithography (VMFCL). The actuator moves the tip to the selected coordinates one by one. For each coordinate, the STM control unit adds a modulation voltage to the dc bias voltage, while the tip hovers over the desired atom. Then, it increases the amplitude of the modulation voltage by a small amount at each sampling time and monitors the tip-sample height. If the control system detects a step change in height that is higher than a predefined threshold  or if the modulation voltage amplitude is increased to a predefined maximum value, then it immediately ramps down the modulation amplitude and moves the tip to the next coordinate.

\begin{figure}[!htb]
	\noindent
	\centering
	\includegraphics[width=1\linewidth]{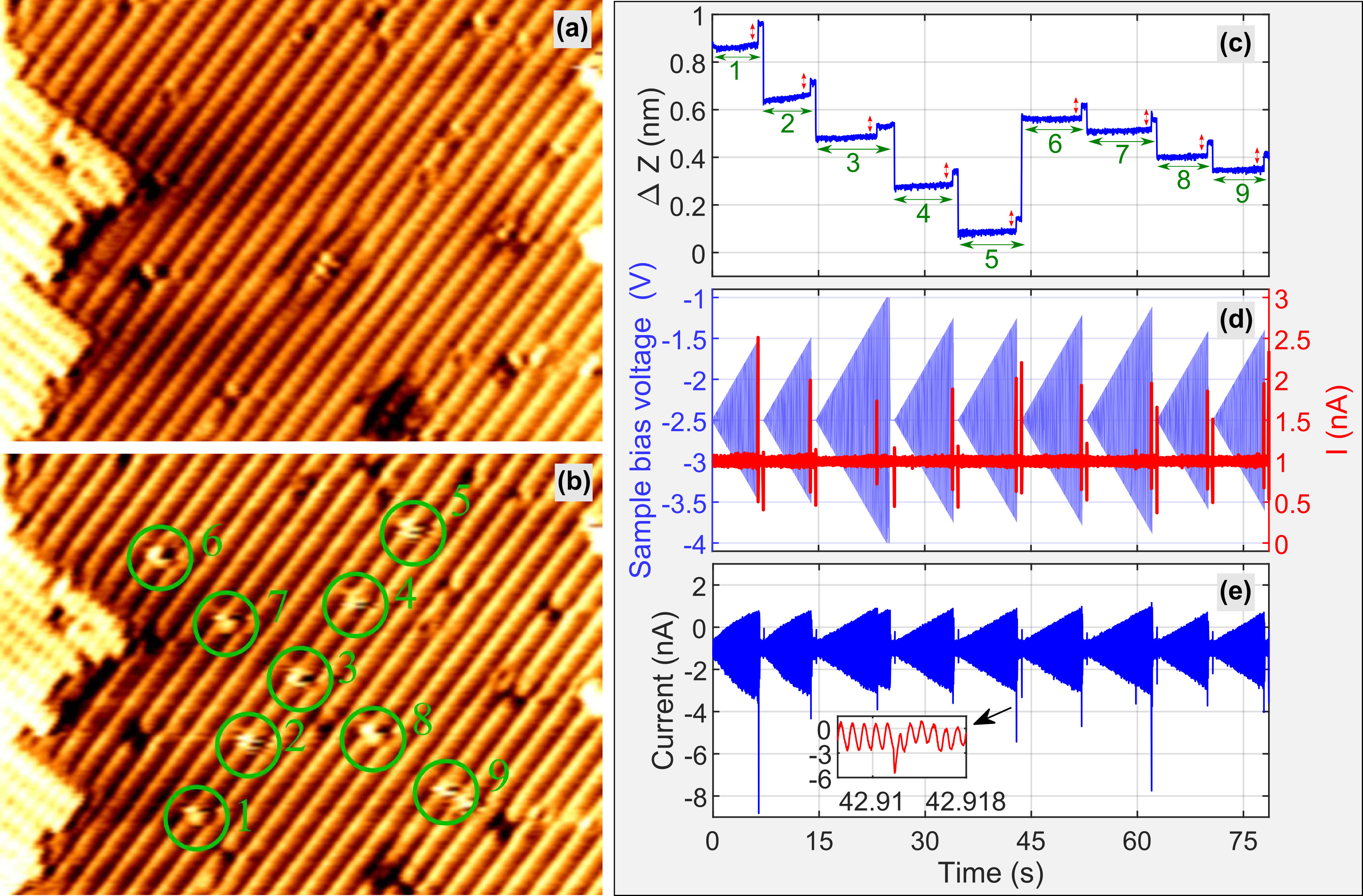}
	\caption{Hydrogen depassivation performed on an H-passivated Si surface using VMFCL. (a) The surface is scanned and nine desorption locations are selected. (b) Hydrogen atoms are removed form the surface one by one, (c) while the tip displacement in z-direction is monitored during the entire process. (d) The modulation amplitude ramps down to zero as soon as a jump [displayed by the red arrows in Figure (c)] higher than the predefined threshold \mbox{(0.3 \AA)} is detected in the z-direction. (e) Adding modulation to the sample induces high frequency oscillations on the current signal, which are filtered out by the notch filters [red curve in Figure (d)]. Figure reprinted with permission from \cite{Hamed_STM_02}, copyright 2020, AIP.}
	\label{fig:VMFCL}
\end{figure}

{Figure~\ref{fig:VMFCL} shows automated removal of H atoms from an H-passivated Si surface using the VMFCL method \cite{Hamed_STM_02}. In this experiment, the sample bias voltage, the setpoint current, and the modulation frequency are \mbox{-2.5\,V}, \mbox{1\,nA}, and \mbox{1\,kHz}, respectively. The modulation amplitude increases at the rate of \mbox{0.15\,V/sec} and is quickly ramped down to zero as it reaches the predefined maximum value of \mbox{1.5\,V} or as a depassivation event is detected. All of the targeted hydrogen atoms are successfully removed from the surface.}
%%%%%%%%%%%%%%%%%%%%%%%%%%%%%%%%%%%%%%%%%%%%%%%%%

%%%%%%%%%%% Conclusion %%%%%%%%%%%%%%%%%%%%%%%%%%
\section{Conclusions}
\label{sec:Conclusions}
In this article, we reviewed some of the recent advances in control of STMs and their implications on imaging, spectroscopy, and atomic-precision lithography performed with this instrument. We analyzed the STM control system in constant current and constant differential conductance modes, detailed a method of designing the PI controller based on an experimentally obtained model of the STM system dynamics, and explained the connection between the dc gain of the STM transfer function and the local barrier height. We reviewed two methods for estimating the LBH variations and described how they may be used to tune the PI controller gains in real time and went on to explain that real-time adaptation of the PI controller gain lowers the risk of a tip-sample crash, thus increasing the tip lifetime. Several spectroscopy methods were reviewed, and the control system impact on the \Ind{} measurements was discussed. We described that incorporating notch filters in the feedback loop greatly enhances the SNR of spectroscopy measurements and explained how ultrafast STS was made possible using this concept to obtain \mbox{I-V} curves in real time. We described limitations of the existing HDL methods and how precise removal of single H atoms was made possible by making small changes to the structure of the STM feedback control system. 

The STM was invented about four decades ago. To this day, it remains the best tool for performing imaging, spectroscopy, and lithography on surfaces with atomic precision. Many of the limitations of this device can be traced back to the lack of performance of its simple feedback control system, which is often tuned by an operator with a limited knowledge of feedback control. Our research demonstrates that understanding the physics of scanning tunneling microscopy and its relationship with the closed-loop dynamics of the system is the key to designing controllers that improve reliability of and enable new or improved functionalities in STM.
%%%%%%%%%%%%%%%%%%%%%%%%%%%%%%%%%%%%%%%%%%%%%%%%%

%%%%%%%%%%% Acknowledgments %%%%%%%%%%%%%%%%%%%%%
\section{Acknowledgments}
\label{sec:Acknowledgments}

This work was supported by the U.S. Department of Energy's Office of Energy Efficiency and Renewable Energy (EERE) under the Advanced Manufacturing Office Award No. DE-EE0008322, and the UT Dallas Center for Atomically Precise Fabrication of Solid-State Quantum Devices.
%%%%%%%%%%%%%%%%%%%%%%%%%%%%%%%%%%%%%%%%%%%%%%%%%

\bibliographystyle{unsrt}
\bibliography{References}

\begin{thebibliography}{10}

\bibitem{wolf_2012}
E.~L. Wolf.
\newblock {\em Principles of Electron Tunneling Spectroscopy: Second Edition}.
\newblock International Series of Monographs on Physics. OUP Oxford, 2012.

\bibitem{Book_Voigtlaender}
B.~Voigtländer.
\newblock {\em {Scanning Probe Microscopy}}.
\newblock Springer Berlin Heidelberg, 2015.

\bibitem{Farid_TCST}
F.~Tajaddodianfar, S.~O.~R. Moheimani, and J.~N. Randall.
\newblock Scanning tunneling microscope control: A self-tuning pi controller based on online local barrier height estimation*.
\newblock {\em IEEE Transactions on Control Systems Technology}, 27(5):2004--2015, 2019.

\bibitem{STS_Feenstra}
R.~M. Feenstra.
\newblock Scanning tunneling spectroscopy.
\newblock {\em Surface Science}, 299-300:965--979, 1994.

\bibitem{Walsh_2009}
M.~A. Walsh and M.~C. Hersam.
\newblock Atomic-scale templates patterned by ultrahigh vacuum scanning tunneling microscopy on silicon.
\newblock {\em Annual Review of Physical Chemistry}, 60(1):193--216, 2009.

\bibitem{Lyding_1994}
J.~W. Lyding, T.‐C. Shen, J.~S. Hubacek, J.~R. Tucker, and G.~C. Abeln.
\newblock Nanoscale patterning and oxidation of {H‐passivated Si(100)‐2\texttimes1} surfaces with an ultrahigh vacuum scanning tunneling microscope.
\newblock {\em Applied Physics Letters}, 64(15):2010--2012, 1994.

\bibitem{Lyding_Abeln_1994}
J.~W. Lyding, G.~C. Abeln, T.‐C. Shen, C.~Wang, and J.~R. Tucker.
\newblock Nanometer scale patterning and oxidation of silicon surfaces with an ultrahigh vacuum scanning tunneling microscope.
\newblock {\em Journal of Vacuum Science \& Technology B: Microelectronics and Nanometer Structures Processing, Measurement, and Phenomena}, 12(6):3735--3740, 1994.

\bibitem{achal_2018_lithography}
R.~Achal, M.~Rashidi, J.~Croshaw, D.~Churchill, M.~Taucer, T.~Huff, M.~Cloutier, J.~Pitters, and R.~A. Wolkow.
\newblock Lithography for robust and editable atomic-scale silicon devices and memories.
\newblock {\em Nature communications}, 9(1):1--8, 2018.

\bibitem{Butcher_2000_lithography}
M.~J. Butcher, F.~H. Jones, and P.~H. Beton.
\newblock Growth and modification of ag islands on hydrogen terminated si(100) surfaces.
\newblock {\em Journal of Vacuum Science \& Technology B: Microelectronics and Nanometer Structures Processing, Measurement, and Phenomena}, 18(1):13--15, 2000.

\bibitem{Adams_1996_lithography}
D.~P. Adams, T.~M. Mayer, and B.~S. Swartzentruber.
\newblock Selective area growth of metal nanostructures.
\newblock {\em Applied Physics Letters}, 68(16):2210--2212, 1996.

\bibitem{Buch_2015_doping}
H.~B\"uch, M.~Fuechsle, W.~Baker, M.~G. House, and M.~Y. Simmons.
\newblock Quantum dot spectroscopy using a single phosphorus donor.
\newblock {\em Phys. Rev. B}, 92:235309, Dec 2015.

\bibitem{Goh_2016_doping}
K.~E.~J. Goh, L.~Oberbeck, M.~Y. Simmons, A.~R. Hamilton, and M.~J. Butcher.
\newblock Influence of doping density on electronic transport in degenerate si:p $\ensuremath{\delta}$-doped layers.
\newblock {\em Phys. Rev. B}, 73:035401, Jan 2006.

\bibitem{LydingScanner}
S.~H. Tessmer, D.~J.~Van Harlingen, and J.~W. Lyding.
\newblock {Integrated cryogenic scanning tunneling microscopy and sample preparation system}.
\newblock {\em Review of Scientific Instruments}, 65(9):2855--2859, 1994.

\bibitem{Kitchin_1986_RMStoDC}
C.~Kitchin and L.~Counts.
\newblock {\em RMS-to-DC Conversion Application Guide: Second Edition}.
\newblock Analog Devices, 1986.

\bibitem{Ando_2001_PeakHold}
T.~Ando, N.~Kodera, E.~Takai, D.~Maruyama, K.~Saito, and A.~Toda.
\newblock A high-speed atomic force microscope for studying biological macromolecules.
\newblock {\em Proceedings of the National Academy of Sciences}, 98(22):12468--12472, 2001.

\bibitem{Ragazzon_2016_PeakDetector}
M.~R.~P. Ragazzon, J.~T. Gravdahl, and A.~J. Fleming.
\newblock On amplitude estimation for high-speed atomic force microscopy.
\newblock In {\em 2016 American Control Conference (ACC)}, pages 2635--2642, 2016.

\bibitem{ruppert_2017_demodulation}
M.~G. Ruppert, D.~M. Harcombe, M.~R.~P. Ragazzon, S.~O.~R. Moheimani, and A.~J. Fleming.
\newblock A review of demodulation techniques for amplitude-modulation atomic force microscopy.
\newblock {\em Beilstein journal of nanotechnology}, 8(1):1407--1426, 2017.

\bibitem{Harcombe_2020_demodulation}
D.~M. Harcombe, M.~G. Ruppert, and A.~J. Fleming.
\newblock A review of demodulation techniques for multifrequency atomic force microscopy.
\newblock {\em Beilstein journal of nanotechnology}, 11(1):76--91, 2020.

\bibitem{Cosens_1934_LockInAmp}
C.~R. Cosens.
\newblock A balance-detector for alternating-current bridges.
\newblock {\em Proceedings of the Physical Society}, 46(6):818--823, 1934.

\bibitem{Michels_1941}
Walter~C. Michels and Norma~L. Curtis.
\newblock A pentode lock‐in amplifier of high frequency selectivity.
\newblock {\em Review of Scientific Instruments}, 12(9):444--447, 1941.

\bibitem{Kokavecz_2006_CoherentDemodulator}
J.~Kokavecz, Z.~T{\'{o}}th, Z.~L. Horv{\'{a}}th, P.~Heszler, and {\'{A}}.~Mechler.
\newblock Novel amplitude and frequency demodulation algorithm for a virtual dynamic atomic force microscope.
\newblock {\em Nanotechnology}, 17(7):S173--S177, 2006.

\bibitem{Abramovitch_2011}
Daniel~Y. Abramovitch.
\newblock Low latency demodulation for atomic force microscopes, part i efficient real-time integration.
\newblock In {\em Proceedings of the 2011 American Control Conference}, pages 2252--2257, 2011.

\bibitem{Ruppert_2016_Kalman}
M.~G. Ruppert, D.~M. Harcombe, and S.~O.~R. Moheimani.
\newblock High-bandwidth demodulation in mf-afm: A kalman filtering approach.
\newblock {\em IEEE/ASME Transactions on Mechatronics}, 21(6):2705--2715, 2016.

\bibitem{Kalman_1960}
R.~E. Kalman.
\newblock {A New Approach to Linear Filtering and Prediction Problems}.
\newblock {\em Journal of Basic Engineering}, 82(1):35--45, 03 1960.

\bibitem{Ragazzon_2018_Lyapunov}
M.~R.~P. Ragazzon, M.~G. Ruppert, D.~M. Harcombe, A.~J. Fleming, and J.~T. Gravdahl.
\newblock Lyapunov estimator for high-speed demodulation in dynamic mode atomic force microscopy.
\newblock {\em IEEE Transactions on Control Systems Technology}, 26(2):765--772, 2018.

\bibitem{Gottlieb_2006}
Alex~D Gottlieb and Lisa Wesoloski.
\newblock Bardeen’s tunnelling theory as applied to scanning tunnelling microscopy: a technical guide to the traditional interpretation.
\newblock {\em Nanotechnology}, 17(8):R57, mar 2006.

\bibitem{STM_Binnig01}
G.~Binnig, H.~Rohrer, Ch. Gerber, and E.~Weibel.
\newblock Tunneling through a controllable vacuum gap.
\newblock {\em Applied Physics Letters}, 40(2):178--180, 1982.

\bibitem{Farid_RSI}
F.~Tajaddodianfar, S.~O.~R. Moheimani, J.~Owen, and J.~N. Randall.
\newblock On the effect of local barrier height in scanning tunneling microscopy: Measurement methods and control implications.
\newblock {\em Review of Scientific Instruments}, 89(1):013701, 2018.

\bibitem{Tajaddodianfar_2017_Control}
F.~Tajaddodianfar, S.~O.~R. Moheimani, J.~Owen, and J.~N. Randall.
\newblock A self-tuning controller for high-performance scanning tunneling microscopy.
\newblock In {\em 2017 IEEE Conference on Control Technology and Applications (CCTA)}, pages 106--110, 2017.

\bibitem{kahn_2016}
Antoine Kahn.
\newblock Fermi level, work function and vacuum level.
\newblock {\em Materials Horizons}, 3(1):7--10, 2016.

\bibitem{fischer_1993}
R~Fischer, S~Schuppler, N~Fischer, Th~Fauster, and W~Steinmann.
\newblock Image states and local work function for ag/pd (111).
\newblock {\em Physical review letters}, 70(5):654, 1993.

\bibitem{jiang_2017}
Yingda Jiang, Jingtai Li, Guirong Su, Nicola Ferri, Wei Liu, and Alexandre Tkatchenko.
\newblock Tuning the work function of stepped metal surfaces by adsorption of organic molecules.
\newblock {\em Journal of Physics: Condensed Matter}, 29(20):204001, 2017.

\bibitem{Alemansour_fastSTS}
H.~Alemansour, S.~O.~R. Moheimani, J.~H.~G. Owen, J.~N. Randall, and E.~Fuchs.
\newblock Ultrafast method for scanning tunneling spectroscopy.
\newblock {\em Journal of Vacuum Science \& Technology B}, 39(4):042802, 2021.

\bibitem{Control_Oliva_1995}
A.~I. Oliva, E.~Anguiano, N.~Denisenko, M.~Aguilar, and J.~L. Peña.
\newblock Analysis of scanning tunneling microscopy feedback system.
\newblock {\em Review of Scientific Instruments}, 66(5):3196--3203, 1995.

\bibitem{Control_Anguiano_1998}
E.~Anguiano, A.~I. Oliva, and M.~Aguilar.
\newblock Optimal conditions for imaging in scanning tunneling microscopy: Theory.
\newblock {\em Review of Scientific Instruments}, 69(11):3867--3874, 1998.

\bibitem{Moheimani_2019_LBHpatent}
S.~O.~R. Moheimani, F.~Tajaddodianfar, E.~Fuchs, J.~Randall, J.~Ballard, and J.~Owen.
\newblock Methods, devices and systems for scanning tunneling microscopy control system design, December~3 2019.
\newblock US Patent 10,495,665.

\bibitem{Lang_1988_GapModulation}
N.~D. Lang.
\newblock Apparent barrier height in scanning tunneling microscopy.
\newblock {\em Phys. Rev. B}, 37:10395--10398, Jun 1988.

\bibitem{Binnig_2000_GapModulation}
G.~Binnig and H.~Rohrer.
\newblock Scanning tunneling microscopy.
\newblock {\em IBM Journal of Research and Development}, 44(1.2):279--293, 2000.

\bibitem{Maeda_2004_GapModulation}
Y.~Maeda, M.~Okumura, S.~Tsubota, M.~Kohyama, and M.~Haruta.
\newblock Local barrier height of au nanoparticles on a tio2(1 1 0)-(1×2) surface.
\newblock {\em Applied Surface Science}, 222(1):409--414, 2004.

\bibitem{STS_Owen}
J.~H.~G. Owen, D.~R. Bowler, C.~M. Goringe, K.~Miki, and G.~A.~D. Briggs.
\newblock {Identification of the Si(001) missing dimer defect structure by low bias voltage STM and LDA modelling}.
\newblock {\em Surface Science}, 341(3):L1042--L1047, 1995.

\bibitem{BiasDependentSTS_01}
R.~J. Hamers.
\newblock {Atomic-resolution surface spectroscopy with the scanning tunneling microscope}.
\newblock {\em Annual Review of Physical Chemistry}, 40(1):531--559, 1989.

\bibitem{BiasDependentSTS_02}
P.~Sutter, P.~Zahl, E.~Sutter, and J.~E. Bernard.
\newblock {Energy-filtered scanning tunneling microscopy using a semiconductor tip}.
\newblock {\em Physical review letters}, 90(16):166101, 2003.

\bibitem{DB_Wolkow01}
H.~Labidi, M.~Taucer, M.~Rashidi, M.~Koleini, L.~Livadaru, J.~Pitters, M.~Cloutier, M.~Salomons, and R.~A. Wolkow.
\newblock {Scanning tunneling spectroscopy reveals a silicon dangling bond charge state transition}.
\newblock {\em New Journal of Physics}, 17(7):073023, jul 2015.

\bibitem{IETS_01}
G.~Binnig, N.~Garcia, and H.~Rohrer.
\newblock {Conductivity sensitivity of inelastic scanning tunneling microscopy}.
\newblock {\em Physical Review B}, 32(2):1336--1338, 1985.

\bibitem{IETS_02}
J.~Lambe and R.~C. Jaklevic.
\newblock {Molecular Vibration Spectra by Inelastic Electron Tunneling}.
\newblock {\em Physical Review}, 165(3):821--832, January 1968.

\bibitem{IETS_03}
A.~S. Hallbäck, N.~Oncel, J.~Huskens, H.~J. Zandvliet, and B.~Poelsema.
\newblock {Inelastic Electron Tunneling Spectroscopy on Decanethiol at Elevated Temperatures}.
\newblock {\em Nano Letters}, 4(12):2393--2395, December 2004.

\bibitem{Hamed_patent}
S.~O.~R. Moheimani and H.~Alemansour.
\newblock Methods and devices configured to operated scanning tunneling microscopes using out-of-bandwidth frequency components added to bias voltage and related software, May~6 2021.
\newblock US Patent App. 17/089,214.

\bibitem{Hamed_STM_03}
H.~Alemansour, S.~O.~R. Moheimani, J.~H.~G. Owen, J.~N. Randall, and E.~Fuchs.
\newblock {High signal-to-noise ratio differential conductance spectroscopy}.
\newblock {\em Journal of Vacuum Science \& Technology B}, 39(1):010601, 2021.

\bibitem{STS_IndividualAtoms}
R.~M. Feenstra, J.~A. Stroscio, J.~Tersoff, and A.~P. Fein.
\newblock {Atom-selective imaging of the GaAs (110) surface}.
\newblock {\em Physical Review Letters}, 58(12):1192, 1987.

\bibitem{bonnell}
D.~Bonnell.
\newblock {\em Scanning Probe Microscopy and Spectroscopy: Theory, Techniques, and Applications}.
\newblock Wiley, 2000.

\bibitem{ZyvexWeb}
\url{https://www.zyvexlabs.com/apm/products/zyvector/}, Online; accessed 02-June-2023.

\bibitem{CITS_time}
A.~Belianinov, P.~Ganesh, W.~Lin, B.~C. Sales, A.~S. Sefat, S.~Jesse, M.~Pan, and S.~V. Kalinin.
\newblock {Research Update: Spatially resolved mapping of electronic structure on atomic level by multivariate statistical analysis}.
\newblock {\em APL Materials}, 2(12):120701, 2014.

\bibitem{Avouris_CPL_1996_HDL}
P.~Avouris, R.~E. Walkup, A.~R. Rossi, T.~C. Shen, G.~C. Abeln, J.~R. Tucker, and J.~W. Lyding.
\newblock {STM}-induced {H} atom desorption from {Si}(100): isotope effects and site selectivity.
\newblock {\em Chemical Physics Letters}, 257(1):148 -- 154, 1996.

\bibitem{Hersam_2000_FCL}
M.~C. Hersam, N.~P. Guisinger, and J.~W. Lyding.
\newblock Silicon-based molecular nanotechnology.
\newblock {\em Nanotechnology}, 11(2):70--76, jun 2000.

\bibitem{Avouris_1996_HDL}
P.~Avouris, R.~E. Walkup, A.~R. Rossi, H.~C. Akpati, P.~Nordlander, T.~C. Shen, G.~C. Abeln, and J.~W. Lyding.
\newblock Breaking individual chemical bonds via {STM}-induced excitations.
\newblock {\em Surface Science}, 363(1):368--377, 1996.
\newblock Dynamical Quantum Processes on Solid Surfaces.

\bibitem{Dagata_1990_HDL}
J.~A. Dagata, J.~Schneir, H.~H. Harary, C.~J. Evans, M.~T. Postek, and J~Bennett.
\newblock Modification of hydrogen-passivated silicon by a scanning tunneling microscope operating in air.
\newblock {\em Applied Physics Letters}, 56(20):2001--2003, 1990.

\bibitem{Ruess_2004_doping}
F.~J. Ruess, L.~Oberbeck, M.~Y. Simmons, K.~E.~J. Goh, A.~R. Hamilton, T.~Hallam, S.~R. Schofield, N.~J. Curson, and R.~G. Clark.
\newblock Toward atomic-scale device fabrication in silicon using scanning probe microscopy.
\newblock {\em Nano Letters}, 4(10):1969--1973, 2004.

\bibitem{RevModPhys.85.961}
Floris~A. Zwanenburg, Andrew~S. Dzurak, Andrea Morello, Michelle~Y. Simmons, Lloyd C.~L. Hollenberg, Gerhard Klimeck, Sven Rogge, Susan~N. Coppersmith, and Mark~A. Eriksson.
\newblock Silicon quantum electronics.
\newblock {\em Rev. Mod. Phys.}, 85:961--1019, Jul 2013.

\bibitem{Soukiassian_2003_APL}
L.~Soukiassian, A.~J. Mayne, M.~Carbone, and G.~Dujardin.
\newblock Atomic-scale desorption of {H} atoms from the $\mathrm{Si}(100)\ensuremath{-}2\ifmmode\times\else\texttimes\fi{}1:\mathrm{H}$ surface: Inelastic electron interactions.
\newblock {\em Phys. Rev. B}, 68:035303, Jul 2003.

\bibitem{Randall_2019_HDL}
J.~N. Randall, J.~H.~G. Owen, J.~Lake, and E.~Fuchs.
\newblock Next generation of extreme-resolution electron beam lithography.
\newblock {\em Journal of Vacuum Science \& Technology B}, 37(6):061605, 2019.

\bibitem{Shen_1995_FEM}
T.~C. Shen, C.~Wang, G.~C. Abeln, J.~R. Tucker, J.~W. Lyding, P.~Avouris, and R.~E. Walkup.
\newblock Atomic-scale desorption through electronic and vibrational excitation mechanisms.
\newblock {\em Science}, 268(5217):1590--1592, 1995.

\bibitem{Stokbro_1998_HDL}
K.~Stokbro, C.~Thirstrup, M.~Sakurai, U.~Quaade, Ben Yu-Kuang Hu, F.~Perez-Murano, and F.~Grey.
\newblock {STM}-induced hydrogen desorption via a hole resonance.
\newblock {\em Phys. Rev. Lett.}, 80:2618--2621, Mar 1998.

\bibitem{Hamed_STM_02}
H.~Alemansour, S.~O.~R. Moheimani, J.~H.~G. Owen, J.~N. Randall, and E.~Fuchs.
\newblock {Controlled removal of hydrogen atoms from H-terminated silicon surfaces}.
\newblock {\em Journal of Vacuum Science {\&} Technology B}, 38(4):040601, July 2020.

\bibitem{Basu_2004_FCL}
R.~Basu, N.~P. Guisinger, M.~E. Greene, and M.~C. Hersam.
\newblock Room temperature nanofabrication of atomically registeredheteromolecular organosilicon nanostructures using multistepfeedback controlled lithography.
\newblock {\em Applied Physics Letters}, 85(13):2619--2621, 2004.

\bibitem{Moller_2017_FCL}
M.~M{\o}ller, S.~P. Jarvis, L.~Gu{\'e}rinet, P.~Sharp, R.~Woolley, P.~Rahe, and P.~Moriarty.
\newblock Automated extraction of single {H} atoms with {STM}: tip state dependency.
\newblock {\em Nanotechnology}, 28(7):075302, 2017.

\bibitem{Hamed_STM_01}
S.~O.~R. Moheimani and H.~Alemansour.
\newblock {A new approach to removing H atoms in hydrogen depassivation lithography}.
\newblock In Eric~M. Panning and Martha~I. Sanchez, editors, {\em Novel Patterning Technologies for Semiconductors, {MEMS}/{NEMS} and {MOEMS} 2020}. {SPIE}, March 2020.

\end{thebibliography}


\begin{thebibliography}{S1}

\bibitem[S1]{SB_Binnig_1982}
G. Binnig, H. Rohrer, C. Gerber, and E. Weibel, ``Tunneling through a controllable vacuum gap,'' \textit{Applied Physics Letters}, vol. 40, no. 2, pp. 178--180, 1982.

\bibitem[S2]{SB_Binnig_1983}
G. Binnig, H. Rohrer, C. Gerber, and E. Weibel, ``7 \ifmmode\times\else\texttimes\fi{} 7 Reconstruction on Si(111) Resolved in Real Space,'' \textit{Physical review letters}, vol. 50, no. 2, pp. 120--123, 1983.

\bibitem[S3]{SB_Binnig_1984}
G. Binnig, N. Garcia, H. Rohrer, J. M. Soler, and F. Flores, ``Electron-metal-surface interaction potential with vacuum tunneling: Observation of the image force,'' \textit{Physical Review B}, vol. 30, no. 8, pp. 4816--4818, 1984.
\end{thebibliography}

\begin{thebibliography}{S2}

\bibitem[S4]{SB_Voogd_2017}
J.M. {de Voogd}, M.A. {van Spronsen}, F.E. Kalff, B.~Bryant, O.~Ostojić,
  A.M.J. {den Haan}, I.M.N. Groot, T.H. Oosterkamp, A.F. Otte, and M.J. Rost.
\newblock Fast and reliable pre-approach for scanning probe microscopes based
  on tip-sample capacitance.
\newblock {\em Ultramicroscopy}, 181:61--69, 2017.

\end{thebibliography}

\begin{thebibliography}{S4}

\bibitem[S5]{SB_Guu_2005}
Y. H. {Guu},
\newblock AFM surface imaging of AISI D2 tool steel machined by the EDM process.
\newblock {\em Applied Surface Science}, 242(3-4), 245-250, 2005.

\bibitem[S6]{SB_Kramer_2003}
S. Kr{\"a}mer,  R. Fuierer, and C. B. Gorman, 
\newblock Scanning probe lithography using self-assembled monolayers.
\newblock {\em Chemical Reviews}, 103(11) 4367-4418, 2003.

\bibitem[S7]{SB_Giessibl_2013}
F. J. {Giessibl},
\newblock Advances in atomic force microscopy.
\newblock {\em Reviews of modern physics}, 75(3), 2013.

\bibitem[S8]{SB_Voigtlander_2015}
B. Voigtl{\"a}nder,
\newblock Scanning probe microscopy.
\newblock {\em Springer}, 27(5), 2271-2278, 2015.

\bibitem[S9]{SB_Coskun_2018}
M. B. Coskun, H. Alemansour, A. G. Fowler, M. Maroufi, and S. O. R. Moheimani, 
\newblock $ Q $ Control of an Active {AFM} Cantilever With Differential Sensing Configuration.
\newblock {\em IEEE Transactions on Control Systems Technology}, 27(5), 2271-2278, 2018.

\end{thebibliography}

\begingroup

\begin{thebibliography}{S4}

\bibitem[S5]{SB04_Simmons_2010}
M.~Fuechsle, S.~Mahapatra, F.A.~Zwanenburg, M.~Friesen, M.A.~Eriksson, and M.Y.~Simmons.
\newblock Spectroscopy of few-electron single-crystal silicon quantum dots.
\newblock {\em Nature Nanotechnology}, 5(7):502--505, 2010.

\bibitem[S6]{SB04_Simmons_2018}
M.A.~Broome, S.K.~Gorman, M.G.~House, S.J.~Hile, J.G.~Keizer, D.~Keith, C.D.~Hill, T.F.~Watson, W.J.~Baker, L.C.L.~Hollenberg, and M.Y.~Simmons.
\newblock Two-electron spin correlations in precision placed donors in silicon.
\newblock {\em Nature communications}, 9(1):1--7, 2018.

\bibitem[S7]{SB04_Simmons_2019}
Y.~He, S.K.~Gorman, D.~Keith, L.~Kranz, J.G.~Keizer, and M.Y.~Simmons.
\newblock A two-qubit gate between phosphorus donor electrons in silicon.
\newblock {\em Nature}, 571(7765):371--375, 2019.

\bibitem[S8]{SB04_Simmons_2017}
T.F.~Watson, B.~Weber, Y.L.~Hsueh, L.C.~Hollenberg, R.~Rahman, and M.Y.~Simmons.
\newblock Atomically engineered electron spin lifetimes of 30 s in silicon.
\newblock {\em Science advances}, 3(3):e1602811, 2017.

\bibitem[S9]{SB04_Simmons_2016}
S.~Shamim, B.~Weber, D.W.~Thompson, M.Y.~Simmons, and A.~Ghosh.
\newblock Ultralow-noise atomic-scale structures for quantum circuitry in
  silicon.
\newblock {\em Nano Letters}, 16(9):5779--5784, 2016.

\end{thebibliography}
\endgroup

\processdelayedfloats % place endfloats here, before the sidebars

\sidebars % changes numbering scheme
%%%%%%%%%%% Sidebar01 %%%%%%%%%%%%%%%%%%%%%%%%%%%
\section[Summary]{Sidebar: Summary}
\label{sidebar:Summary}

\setcounter{equation}{0}
\renewcommand{\theequation}{S\arabic{equation}}
\setcounter{table}{0}
\renewcommand{\thetable}{S\arabic{table}}
\setcounter{figure}{0}
\renewcommand{\thefigure}{S\arabic{figure}}

The invention of scanning tunneling microscope (STM) dates back to the work of Binnig and Rohrer in the early 1980s \cite{SB_Binnig_1982, SB_Binnig_1983, SB_Binnig_1984}, whose seminal contribution was rewarded by the 1986 Nobel Prize in Physics “for the design of the scanning tunneling microscope.” Forty years later, the STM remains the best existing tool  for studying electronic, chemical, and physical properties of conducting and semiconducting surfaces with atomic precision. It has opened entirely new fields of research, enabling scientists to gain invaluable insight into properties and structure of matter at the atomic scale. Recent breakthroughs in STM-based automated hydrogen depassivation lithography (HDL) on silicon have resulted in the STM being considered a viable tool for fabrication of error-free silicon-based quantum-electronic devices. Despite the STM's unique ability to interrogate and manipulate matter with atomic precision, it remains a challenging tool to use. It turns out that many issues can be traced back to the STM's feedback control system, which has remained essentially unchanged since its invention about 40 years ago. This article explains the role of feedback control system of the STM and reviews some of the recent progress made possible in imaging, spectroscopy, and lithography by making appropriate changes to the STM's feedback control loop. We believe that the full potential of the STM is yet to be realized, and the key to new innovations will be the application of  advanced model-based control and estimation techniques to this system.

\clearpage
\newpage
%%%%%%%%%%%%%%%%%%%%%%%%%%%%%%%%%%%%%%%%%%%%%%%%%

%%%%%%%%%%% Sidebar02 %%%%%%%%%%%%%%%%%%%%%%%%%%%
\section[Automated Approach of the Tip to the Sample Surface]
{Sidebar: Automated Approach of the Tip to the Sample Surface}
\label{sidebar:automated_approach}

\setcounter{equation}{0}
\renewcommand{\theequation}{S\arabic{equation}}
\setcounter{table}{0}
\renewcommand{\thetable}{S\arabic{table}}
\setcounter{figure}{0}
\renewcommand{\thefigure}{S\arabic{figure}}

The STM cannot detect the surface until the tip is within a few {\AA}ngstr{\"o}m distance of the surface. As a result, the tip may crash into the surface if it is moved toward it at a rate faster than the current sensing bandwidth. It is crucial to have an efficient method to bring the tip into the tunneling range in a reasonable amount of time. The following approach is commonly employed for this purpose \cite{SB_Voogd_2017}. 

\begin{figure}[!htb]
    \centering  
	\includegraphics[width=0.8\linewidth]{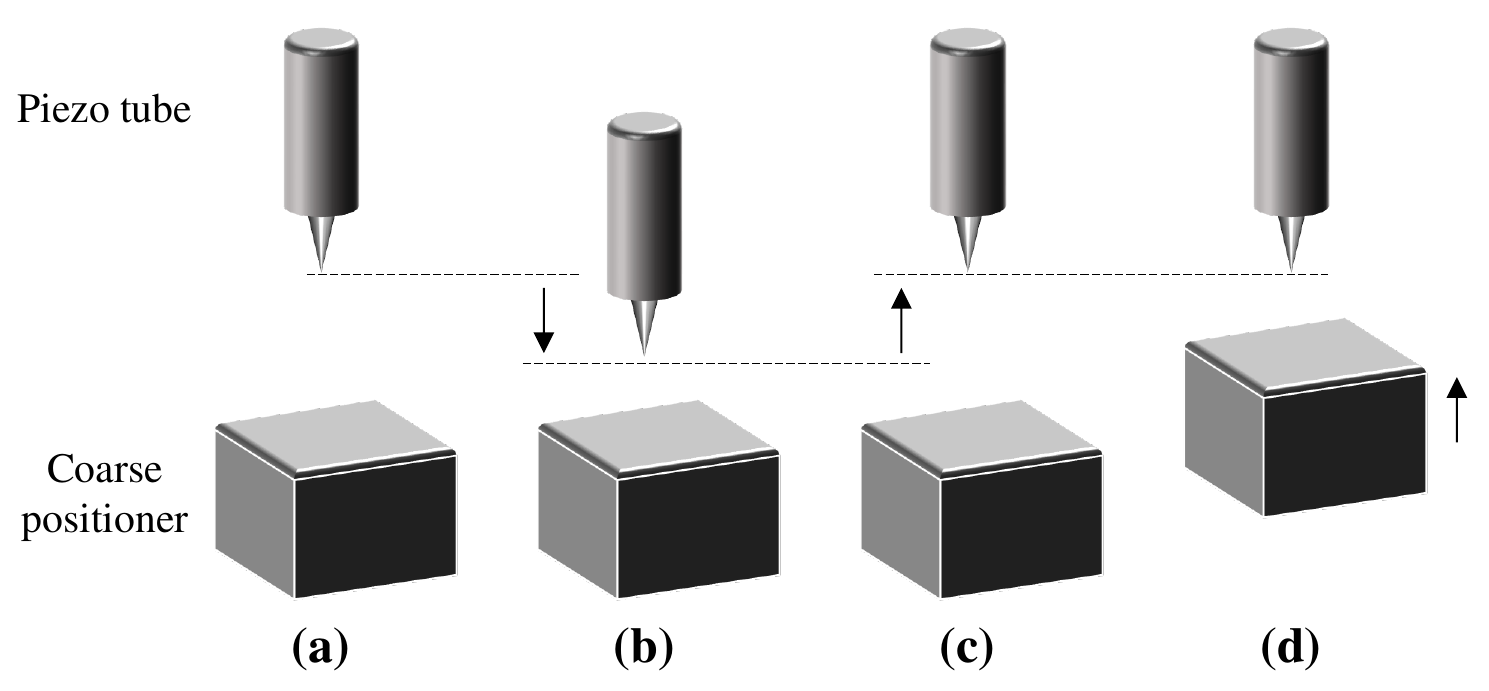}
	\caption{(a) The tip is moved downward a little more than a single step size of the coarse positioner. During this step, the downward motion is stopped whenever the tunneling current is detected. (b) The tip is reached to the end of its extension range or to the point where the tunneling current is detected. (c) The tip is retracted to the original position, knowing that the tunneling current is not detectable within the  extension range for the fine positioner. (d) The sample is moved up one step by the coarse positioner, and the whole process is repeated.
	\label{fig_sb:approach}}
\end{figure}

\begin{enumerate}
    \item Manual positioning: Initially there is a wide gap between the tip and the sample. The user examines this gap with a camera and reduces it by manually moving the sample (or the tip) using the coarse positioner. Since the displacement range of the piezoelectric tube actuator is only a few nanometers, another step is also required to further reduce the gap.
    \item Coarse positioning: This is an automated high-sensitivity stepping procedure. In this step, the piezoelectric tube is extended to its full range, while the tip-sample gap is reduced by the coarse positioner. The tunneling current is monitored by the control system. The piezoelectric tube will retract as soon as the tunneling current is detected.
    \item Fine positioning: In this step, the piezoelectric tube scanner is brought within the desired operating range. This step is schematically shown in Figure~\ref{fig_sb:approach}. Initially, the fine positioner is slowly extended moving the tip towards the sample. The tip movement is stopped as soon as the tunneling current is detected. If the fine positioner is within its desired extension range, the automated positioning process is terminated. Otherwise, the piezoelectric tube is fully retracted. Then, the coarse positioner moves the sample one step towards the tip and this step is repeated.
\end{enumerate}

\newpage
\processdelayedfloats % place sidebar endfloats here
\clearpage
%%%%%%%%%%%%%%%%%%%%%%%%%%%%%%%%%%%%%%%%%%%%%%%%%

%%%%%%%%%%% Sidebar03 %%%%%%%%%%%%%%%%%%%%%%%%%%%
\section[What is Quantum Tunnelling?]
{Sidebar: What is Quantum Tunnelling?}
\label{sidebar:quantum_tunneling}

\setcounter{equation}{0}
\renewcommand{\theequation}{S\arabic{equation}}
\setcounter{table}{0}
\renewcommand{\thetable}{S\arabic{table}}
\setcounter{figure}{1}
\renewcommand{\thefigure}{S\arabic{figure}}

Quantum theory states that an electron can only exist at specific discrete energy levels. Furthermore, each energy state cannot have more than one electron. Energy states are occupied from the lowest to the highest levels, and the top energy level is called the Fermi level. The difference between the Fermi energy and the vacuum energy is called the work function $\phi$. In metals, energy states are very close, and there are many electrons in a tiny amount of material \cite{Book_Voigtlaender}. The density of states $\rho (E)$  is a distribution function that represents the number of states within a finite energy range between $E$ and $E+d\,E$.

\begin{figure}[!htb]
    \centering  
	\includegraphics[width=0.6\linewidth]{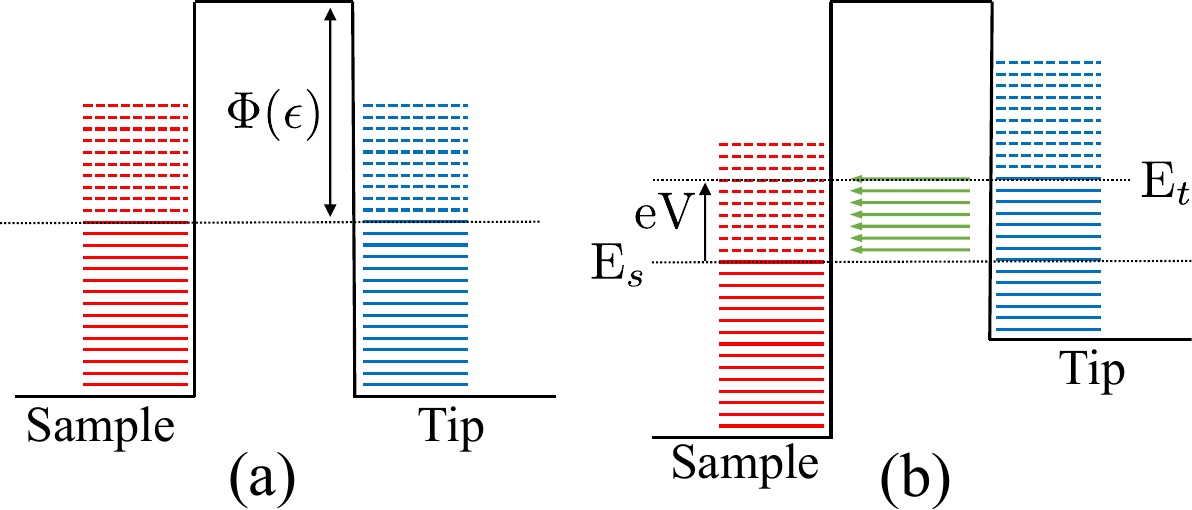}
	\caption{Energy diagram of tip-sample junction when (a) they are at the same potential, and (b) the sample is at a positive potential relative to the tip.
	\label{fig_sb:TunnelingSch}}
\end{figure}

The vacuum or air gap ($\delta$) between the the sample and the tip functions as a barrier for electrons. This can be modeled by a square potential barrier schematically shown in Figure~\ref{fig_sb:TunnelingSch}. According to the quantum theory, for a thin gap, there is a non-zero probability that electrons pass through the barrier and find their way to the other side. At a zero-bias voltage, the net charge passing through the tip-sample gap is zero since the Fermi level of both the sample and the tip is the same (that is, the gradient is zero). Applying a positive bias voltage to the sample decreases energy of its states relative to the tip. Therefore, tunneling can happen in the energy states ranging between $E_s$ to $E_t$. The total current is then proportional to the number of empty states of the sample ($\rho_{s}(\epsilon)$) times the number of filled states of the tip ($\rho_{t}(\epsilon - eV)$) within the energy range between $E_s$ to $E_t$. 

\newpage
\processdelayedfloats % place sidebar endfloats here
\clearpage
%%%%%%%%%%%%%%%%%%%%%%%%%%%%%%%%%%%%%%%%%%%%%%%%%

%%%%%%%%%%% Sidebar04 %%%%%%%%%%%%%%%%%%%%%%%%%%%
\section[How does the Atomic Force Microscope Work?]
{Sidebar: How does the Atomic Force Microscope Work?}
\label{sidebar:AFM}

\setcounter{equation}{0}
\renewcommand{\theequation}{S\arabic{equation}}
\setcounter{table}{0}
\renewcommand{\thetable}{S\arabic{table}}
\setcounter{figure}{0}
\renewcommand{\thefigure}{S\arabic{figure}}

The atomic force microscope (AFM) is a mechatronic microscope that can provide sub-nanometer resolution image of surfaces \cite{SB_Guu_2005, SB_Kramer_2003}. STM and AFM both belong to the family of scanning probe microscopes. However, unlike the STM, the AFM can be used to image conducting and nonconducting surfaces. In addition, it can be operated in various environments and does not require an ultrahigh vacuum. The underlying physics of the AFM is  quite different from the STM's. However, they do share similar features, for example, the z-axis controller, lateral scan patterns, and some key building blocks. Thus, the rich literature in AFM control can inspire further development of the STM control techniques \cite{SB_Giessibl_2013}. 

A microcantilever with a sharp tip is the essence of an AFM. Depending on the imaging mode, the tip interacts with the sample in contact, intermittent contact, or non-contact mode. These interactions can be quantified by cantilever bending. Optical beam deflection (OBD) is the most widely used approach to measure the cantilever deflection. In this method, a laser beam is emitted on the cantilever by a laser diode, and the reflected beam is tracked by segmented photodiodes. 

In static AFM, the tip-sample force on the cantilever is counterbalanced by the cantilever bending while the probe scans the surface (for example, in a raster pattern). Constant force and constant height modes are two aspects of this technique \cite{SB_Voigtlander_2015}.

In constant force mode, surface topographic features are mapped by adjusting the cantilever height in a feedback loop such that the cantilever bending (that is, tip-sample force) is kept constant. The control effort is then interpreted as the topography.
In constant height mode, the z-axis is operated in  open loop, starting with an initially preset cantilever deflection.
The cantilever deflections during a scan represent the topography of the surface. Although higher scan speeds can be achieved in this mode, it is not a commonly used method.

In dynamic mode atomic force microscopy, the cantilever is  oscillated at a frequency near its resonance. The displacement of the cantilever is conventionally measured using the OBD method. Tapping mode and frequency modulation (FM) mode are most widely used instances of dynamic AFM.

In tapping mode, the signal detected by the photodiode is demodulated by a lock-in amplifier to determine amplitude and phase of oscillations. During a scan, the tip periodically taps the surface. Tip-sample interactions alter the resonance frequency of the cantilever, changing the oscillation amplitude. A feedback controller moves the cantilever in z-direction to adjust the amplitude of oscillations to the setpoint value. A topographic image of surface is obtained by plotting the control signal against the x- and y-position data \cite{SB_Coskun_2018}.

In frequency modulation mode, the probe is self excited to enable a variable oscillation frequency. Due to the tip-sample interactions during a scan, the resonance frequency shifts, which is then followed by the cantilever oscillation frequency. In this case, the resonance frequency shift ($\Delta \omega$) is measured by a phase-locked loop (PPL)  and utilized to adjust the tip-sample distance in a feedback control loop \cite{SB_Voigtlander_2015}.

\newpage
\processdelayedfloats % place sidebar endfloats here
\clearpage
%%%%%%%%%%%%%%%%%%%%%%%%%%%%%%%%%%%%%%%%%%%%%%%%%

%%%%%%%%%%% Sidebar05 %%%%%%%%%%%%%%%%%%%%%%%%%%%
\section[Atomically Precise Manufacturing for Si Quantum Computing]
{Sidebar: Atomically Precise Manufacturing for Si Quantum Computing}
\label{sidebar:APM}

\setcounter{equation}{0}
\renewcommand{\theequation}{S\arabic{equation}}
\setcounter{table}{0}
\renewcommand{\thetable}{S\arabic{table}}
\setcounter{figure}{2}
\renewcommand{\thefigure}{S\arabic{figure}}

The STM can be used to position single dopant atoms in a Si crystal with atomic precision. This is made possible by creating reactive sites on a H-passivated silicon surface with an STM. Dopant precursor molecules (for example, PH$_3$) selectively react with the exposed Si atoms and decompose into isolated dopants \cite{SB04_Simmons_2010}. These dopant atoms can function as qubits. Fabrication of the first two-qubit gate, which is the central component of quantum computers, is reported in \cite{SB04_Simmons_2018, SB04_Simmons_2019}.

\begin{figure}[!htb]
    \centering  
	\includegraphics[width=1\linewidth]{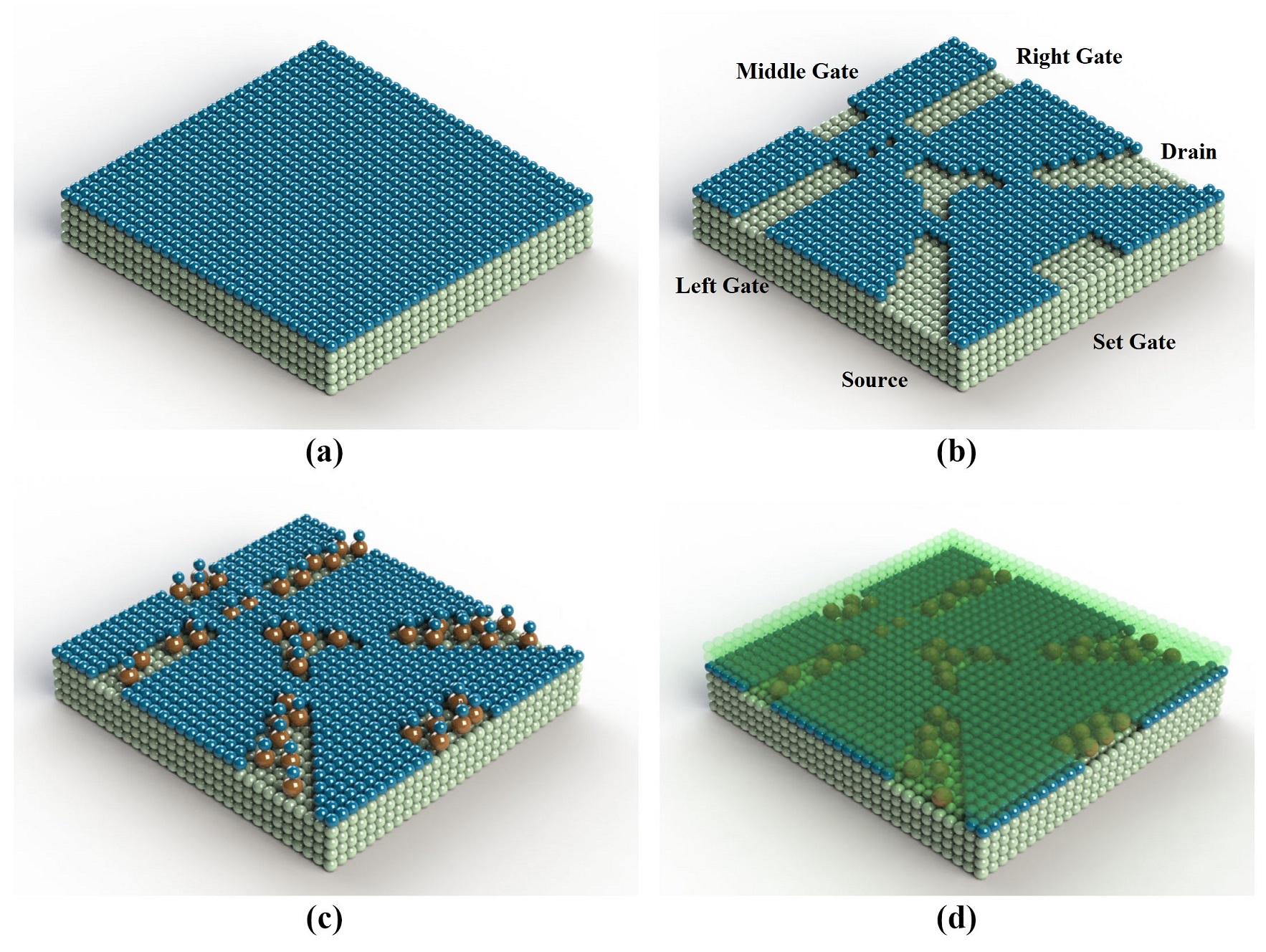}
	\caption{Atomically precise manufacturing of two donor qubits              \cite{SB04_Simmons_2019}. (a) An H-terminated silicon surface is prepared in a UHV STM chamber. (b) The device pattern is written with the STM tip in the high-voltage field emission lithography mode for patterning large areas, as well as the low-voltage atomically precise lithography mode where single hydrogen atoms are depassivated. (c) Doping precursor molecules adsorb and incorporate on the exposed reactive silicon pattern. (d) The incorporated phosphorus atoms are buried by growing an epitaxial Si film on them to protect the device form oxidation.
	\label{fig_sb:APM}}
\end{figure}

The device fabrication method is described in Figure~\ref{fig_sb:APM}. A p-type silicon substrate is passivated with a single layer of hydrogen atoms. The STM tip is then used to make a lithographic mask of the device. Subsequently, the substrate is exposed to gaseous PH$_3$ precursor. The H atoms on the surface act as a resist and limit the possible absorption locations of PH$_3$ to the exposed silicon pattern. Then, a few nanometers of silicon are grown on the device using molecular beam epitaxy. After the burial, prepatterned alignment marks are used to locate the device on the substrate, so that the ohmic contacts can be made.
Researchers have been able to demonstrate exquisite control over high-fidelity sequential readouts of two electron spin qubits  \cite{SB04_Simmons_2017} and reported fabrication of atomic-scale ultrathin Si:P nanowires with the lowest electrical noise to connect to those qubits \cite{SB04_Simmons_2016}.

\newpage
\processdelayedfloats % place sidebar endfloats here
\clearpage
%%%%%%%%%%%%%%%%%%%%%%%%%%%%%%%%%%%%%%%%%%%%%%%%%

%%%%%%%%%%% Bigraphy %%%%%%%%%%%%%%%%%%%%%%%%%%%%
\section{Author Biography}
\label{sec:biography}

\begin{wrapfigure}{l}{0.25\textwidth}
\vspace{-30pt}
    \begin{center}
        \includegraphics[width=0.25\textwidth]{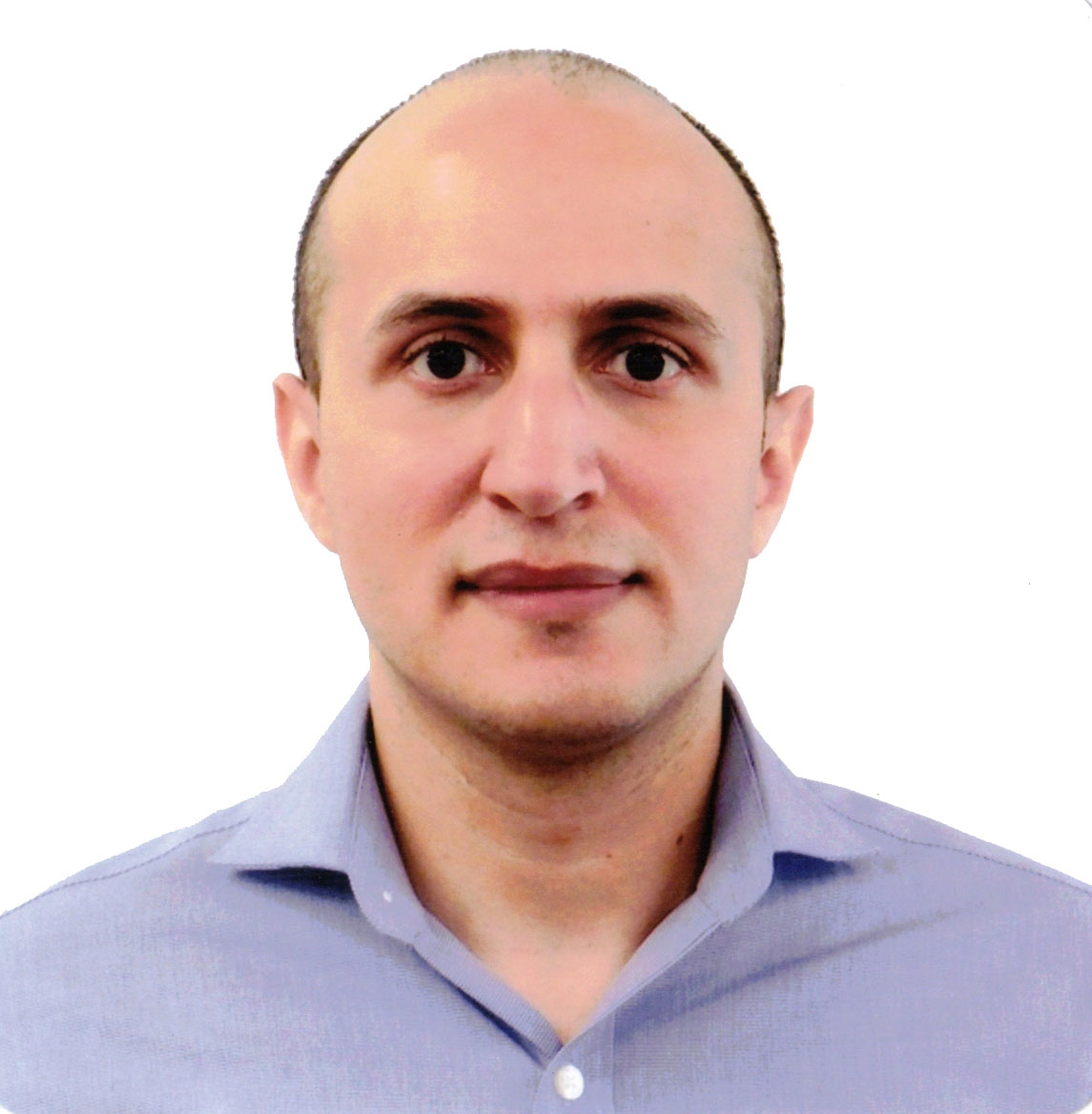}
    \end{center}
\vspace{-30pt}
\end{wrapfigure} \noindent Hamed Alemansour was a PhD student in the mechanical engineering department of the Erik Jonsson School of Engineering and Computer Science at the University of Texas at Dallas (UTD). During his 6 years with UTD, he specialized in the field of ultra-high precision mechatronic systems. His research has ranged across different areas including applications of control in atomically precise manufacturing with scanning tunneling microscope (STM) as well as modeling and control of atomic force microscopes (AFM) and microelectromechanical systems (MEMS). Hamed completed his MSc in mechanical engineering from the Sharif University of Technology, 2013, and his BSc in the same field from the Shiraz University, 2011. He currently holds the position of AFM product scientist in Quantum Design, Inc. San Diego, CA.
\vspace{30pt}

\begin{wrapfigure}{l}{0.40\textwidth}
\vspace{-30pt}
    \begin{center}
        \includegraphics[width=0.40\textwidth]{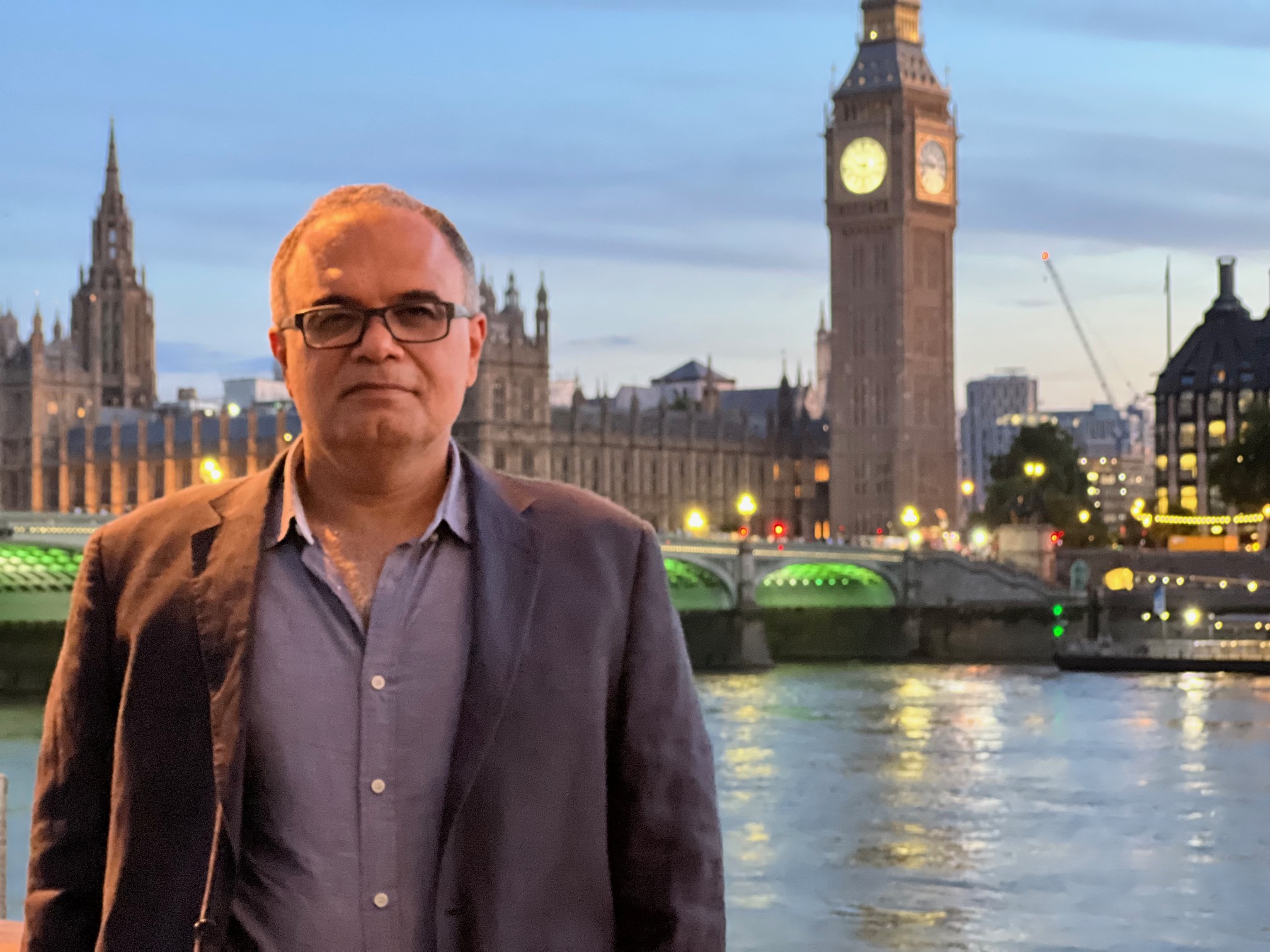}
    \end{center}
\vspace{-30pt}
\end{wrapfigure}
\noindent Reza Moheimani is a professor and holds the James Von Ehr Distinguished Chair in Science and Technology in the Department of Systems Engineering at the University of Texas at Dallas with appointments in Electrical and Computer Engineering and Mechanical Engineering Departments. He is the founding Director of UTD Center for Atomically Precise Fabrication of Solid-State Quantum Devices and founder and Director of Laboratory for Dynamics and Control of Nanosystems. He is a past Editor-in-Chief of Mechatronics (2016-2021), and a past associate editor of IEEE Transactions on Control Systems Technology, IEEE Transactions on Mechatronics and Control Engineering Practice. He received the Industrial achievement Award (IFAC, 2023), Nyquist Lecturer Award (ASME DSCD, 2022), Charles Stark Draper Innovative Practice Award (ASME DSCD, 2020), Nathaniel B. Nichols Medal (IFAC, 2014), IEEE Control Systems Technology Award (IEEE CSS, 2009) and IEEE Transactions on Control Systems Technology Outstanding Paper Award (IEEE CSS, 2007 and 2018). He is a Fellow of IEEE, IFAC, ASME, and Institute of Physics (UK). Moheimani received the Ph.D. degree in Electrical Engineering from University of New South Wales, Australia in 1996. His current research interests include applications of control and estimation in high-precision mechatronic systems, high-speed scanning probe microscopy and atomically precise manufacturing. 
%%%%%%%%%%%%%%%%%%%%%%%%%%%%%%%%%%%%%%%%%%%%%%%%%

\end{document}